\documentclass[%
 reprint,
 amsmath,amssymb,
 aps,
]{revtex4-2}
\usepackage{graphicx}
\usepackage{dcolumn}
\usepackage{bm}
\usepackage{subcaption}
\usepackage[T1]{fontenc}
\usepackage[export]{adjustbox}
\usepackage{color}
\usepackage{hyperref}
\usepackage[mathlines]{lineno}

\begin{document}
\preprint{APS/123-QED}
\title{Investigating the relation between elastic and relaxation properties of dry, frictional granular media during shear deformation}

\author{Aurélien Rigotti}%
\email{aurelien.rigotti@univ-grenoble-alpes.fr}
\affiliation{%
 Univ. Grenoble Alpes, Univ. Savoie Mont Blanc, CNRS, IRD, Grenoble INP, ISTerre, 38000 Grenoble, France}

\author{Véronique Dansereau}%
\email{veronique.dansereau@univ-grenoble-alpes.fr}
\affiliation{%
 Univ. Grenoble Alpes, Univ. Savoie Mont Blanc, CNRS, IRD, Grenoble INP, ISTerre, 38000 Grenoble, France\\
  \altaffiliation[Also at ]{Univ. Grenoble Alpes, CNRS, IRD, Grenoble INP, IGE, 38000 Grenoble, France}}%
\author{Jérôme Weiss}%
 \email{jerome.weiss@univ-grenoble-alpes.fr}
\affiliation{%
 Univ. Grenoble Alpes, Univ. Savoie Mont Blanc, CNRS, IRD, Grenoble INP, ISTerre, 38000 Grenoble, France\\}

\begin{abstract}
\begin{quote}
 Using discrete element simulations based on molecular dynamics, we investigate the mechanical behavior of sheared, dry, frictional granular media in the "dense" and "critical" regimes. We find that this behavior is partitioned between transient stages and a final stationary stage. While the later is macroscopically consistent with the predictions of the viscous, $\mu(I)$ rheology, both the macroscopic behavior during the transient stages and the overall microscopic behavior suggest a more complex picture. Indeed, the simulated granular medium exhibits a finite elastic stiffness throughout its entire shear deformation history, although topological rearrangements of the grains at the microscale translate into a partial degradation of this stiffness, which can be interpreted as a form of elastic damage. The relaxation of stresses follows a compressed exponential, also highlighting the role of elastic interactions in the medium, with residual stresses that depend on the level of elastic damage. The established relations between elastic and relaxation properties point to a complex rheology, characterized by a damage-dependent transition between a visco-elasto-plastic and a viscous behavior.
\end{quote}
\end{abstract}

\maketitle


\section{\label{sec:intro} Introduction}
Granular media such as sand, rice, coffee and powders are encountered in many natural and industrial contexts. They have been extensively studied, thereby fostering a good understanding of their mechanical behavior \cite{LiuNagel1998, Andreotti2013}. These materials are complex in the sense that they exhibit a \textit{jamming transition} \cite{LiuNagel1998,liu2010jamming}. That is, they are characterized by a solid-like behavior in the "jammed" state, with quantifiable elastic properties \cite{Makse2004,Digby1981,Walton1987,Andreotti2013,Johnson1998,Johnson2000,Khidas2010,Rainone2015,Maloney2004,Ishima2020,Karimi2019}, and a fluid-like behavior \cite{LiuNagel1998,GDRMidi2004,daCruz2005,Pouliquen2005,Pouliquen2006,Jop2006,Olsson2007,Olsson2012,Forterre2008} in the "un-jammed" state, in which grain collisions and friction give rise to an effective viscous dissipation. Numerous studies have attempted to quantify the mechanical behavior of granular media and their associated physical properties in both states. \newline

In the case of the jammed state, the Effective Medium Theory (EMT) was developed to predict the non-linear elastic behavior of athermal, 3D granular media with Hertzian particle contacts \cite{Mindlin1949,Mindlin1957,Walton1987}. Within this framework, the bulk and shear elastic modulii, $K$ and $G$, scale with the applied pressure as $K(P) \sim G(P) \sim P^{\alpha}$ where alpha is a positive scaling exponent \cite{Goddard1990}. However, even if non-linear elastic contacts are a good approximation for an experimental granular medium, the underlying assumption of a uniform strain field at all scales \cite{Makse2004} is in general not met. Therefore, in practice, granular media do not follow the EMT power law \cite{Johnson2000} and exhibit complex dependencies on various variables. 
Indeed, a change of exponent $\alpha$ of the EMT scaling when the pressure increases beyond a certain threshold has been noted by Makse et al. \cite{Makse2004}. Elastic properties were shown to depend on the inter-particles friction \cite{Makse2004,Petit2018,Ishima2020}. Variations of the grain-grain friction coefficient, which drive the evolution of porosity ($1 - \phi$, with $\phi$ the packing fraction), were also shown to impact the bulk modulus \cite{Petit2018}. This modulus was moreover observed to decay with increasing polydispersity, monodisperse assemblies being the most rigid ones \cite{Petit2018}. Furthermore, the elastic properties of an amorphous medium are highly sensitive to the preparation of the granular assembly. An initial anisotropy in the preparation has indeed proved to lead to a strong anisotropy of the measured elastic modulii, an effect not taken into account by the EMT \cite{Johnson1998,Khidas2010}. Finally, the shear modulus of amorphous media was also observed to decay drastically in samples subjected to a too brutal precompression \cite{Rainone2015}. 

All of this shows that the elastic properties of a granular medium strongly depend on its topological configuration. Indeed, when a such medium is subjected to external forces and/or to macroscopic deformation, topological rearrangements occur at the grain scale to accommodate deformations and stresses \cite{Radjai2002}. Therefore, it is expected that elastic properties evolve throughout the deformation.
Some authors studied the evolution of elastic properties with respect to the applied (compressive) strain in granular media undergoing small deformations \cite{Karimi2019,Maloney2004}. Using molecular dynamics simulations of a dry frictional granular material, Karimi et al. \cite{Karimi2019} highlighted a correlation between its average coordination number, $Z$, defined as the average number of contacts between each grain and its neighbors, and the shear modulus, $G$. Moreover, they showed that Kachanov's definition of damage in continuum solids \cite{Kachanov1986,Kattan2012} can be extended in the context of granular media when considering the degradation of $G$ that results from topological rearrangements. This phenomenon can be monitored through non-affine displacements of the grains \cite{Maloney2004}. Nevertheless, a complete quantification of such "granular damage" is, to our knowledge, still missing and thus constitutes an interesting topic.

In the case of the unjammed state, that is, below a certain pressure and packing fraction that characterize the \textit{jamming transition} \cite{LiuNagel1998,liu2010jamming}, granular media behave as fluids.
In the general case of dry, frictional granular media, the framework that is perhaps the most widely used to describe the stress-strain relationship in this regime is the $\mu(I)$ rheology, which has been formulated on an empirical and dimensional analysis basis \cite{GDRMidi2004,daCruz2005,Jop2006,Pouliquen2005,Forterre2008}. It combines the following function for the packing fraction 
\begin{equation}
\phi = \phi_{max} - (\phi_{max} - \phi_{min}) I,
    \label{eq:phi_mu_I}
\end{equation}
where $\phi_{max}$ and $\phi_{min}$ are respectively its maximum (or critical) and minimum value, 
and a friction law for the macroscopic shear stress, $\tau$, such that
\begin{equation}
\tau = \mu(I) P,
    \label{eq:tau_P}
\end{equation}
where $\mu(I)$ is a macroscopic friction coefficient. This coefficient is a function of a dimensionless inertial number, $I = \frac{\dot{\gamma} \cdot \bar{d}}{\sqrt{P / \rho_g}}$, which interprets as the ratio of two timescales: the time associated with the confinement of the grains, $d \sqrt{\frac{\rho_g}{P}}$, with $\bar{d}$ the average diameter and $\rho_g$ the density of the grains and $P$ the confinement pressure, and the time associated with the deformation of the media, $\frac{1}{\dot{\gamma}}$ \cite{GDRMidi2004}.

Generalizing this framework in tensorial form to represent the flow of dry granular media involves introducing an effective viscosity for the material, $\eta$, which depends on both the shear rate and pressure, such that 
\begin{equation}
\tau_{ij} = \eta(I,P) \dot{\epsilon}_{ij}
\label{eq:mu_of_I}
\end{equation}
where $\tau_{ij}$ is the deviatoric stress tensor, $\dot{\epsilon_{ij}}$ is the strain rate tensor and $\eta(I, P) = \mu(I) \frac{P}{|\dot{\gamma}|}$, where the denominator represents the second invariant of the strain rate tensor \cite{Jop2006,Pouliquen2006}. 
\begin{figure}[t]
    \centering
    \includegraphics[width=90mm,right]{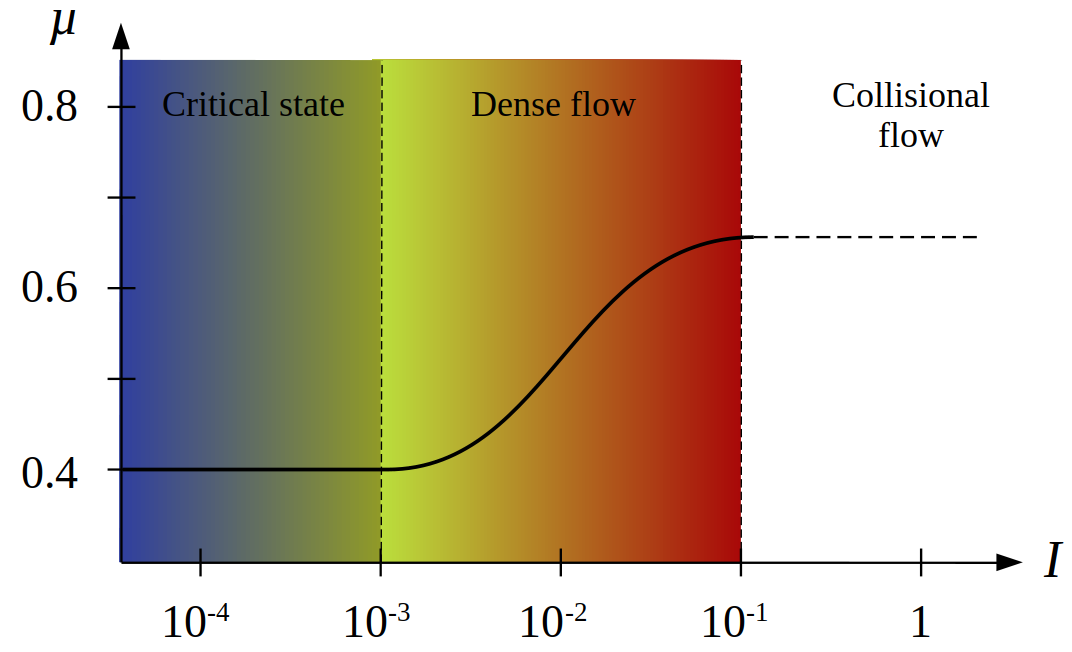}
    \caption{Inertial regimes as a function of the inertial number, with the $\mu(I)$ prediction (black line) for the macroscopic friction coefficient: $\mu(I) = \mu_1 + \frac{\mu_2 - \mu_1}{\frac{I_0}{I} +1}$, where $\mu_1$ is an asymptotic minimum value of $\mu$ when $I \to 0$ and $\mu_2$ an asymptotic maximum value of $\mu$ for $I \gg 10^{-1}$ and $I_0$ is a material-dependent constant \cite{GDRMidi2004,daCruz2005,Jop2006,Pouliquen2005,Forterre2008}.
}
\label{fig:inertial_regimes}
\end{figure}

Three regimes emerge from the dimensional analysis of this rheology, as sketched on FIG. (\ref{fig:inertial_regimes}): a collisional regime for dilute granular media and large inertial numbers, for which the macroscopic friction coefficient $\mu$ tends to a maximum asymptotic value, an intermediate, dense flow regime where the deformation is assumed to be uniform \cite{daCruz2005} and therefore a local rheology relevant, and a critical state at low inertial numbers, for which $\mu$ tends to a minimum asymptotic value.
The dense flow regime has been widely investigated and the $\mu(I)$ rheology well-established in the case of idealised (i.e., low polydispersity and perfectly round-shaped grains) athermal, dry, frictional granular media (e.g. \cite{GDRMidi2004}) and suspensions (using a similar, so-called $\mu(J)$ rheology, \cite{Bonn2017,Guazzelli2018}). In this regime, the inertial number $I$ was also shown to govern the flow properties (velocity, dilatancy, effective friction).
In comparison, the critical regime remains poorly understood. Indeed, when the shear rate vanishes and therefore $I \to 0$, the $\mu(I)$ rheology only predicts that the effective viscosity of the medium diverges to ensure that a yield criterion exists \cite{Jop2006}. The form of this divergence has been studied in frictionless granular media, for $I \to 0$ and low pressures, by Olsson and Teitel \cite{Olsson2007,Olsson2012}, who proposed that the effective viscosity scales as $\eta \sim (\phi - \phi_{max})^{-\beta}$ where $\beta > 1$ and therefore diverges as the packing fraction increases toward the critical packing fraction $\phi_{max}$. 

Yet, when the shear deformation applied to a dry, frictional granular medium is suspended, force chains and internal stresses dissipate. This dissipation can be divided into two stages: an initially fast stage and a later, logarithmic-like and much slower stage \cite{Hartley2003,Miksic2013}. This suggests that, in this solid state with a macroscopic $I=0$, dissipation mechanisms leading to a viscous-like stress relaxation are still at play locally. Interestingly, when the medium is initially compressed instead of sheared, such relaxation does not occur, thereby suggesting a seminal role of shearing in this process \cite{Hartley2003}. Miksic and Alava \cite{Miksic2013} considered that such dissipation can result from both (i) delayed topological grain rearrangements and (ii) a slow, logarithmic "aging" of contact between particles similar to the aging law formulated for friction \cite{Dieterich1979,Brechet1994}. In any case, the time dependence of this stress relaxation in granular media strongly differs from the exponential relaxation associated with classical continuum visco-elastic models.

In general, the structural or stress relaxation (both being likely related \cite{Cipelletti2005}) in amorphous media such as glasses or soft matter \cite{Xia2001,Cipelletti2000,Cipelletti2005} far from equilibrium does not follow the classical exponential form. Instead, the following form has been proposed:
\begin{equation}
    f(t) \sim exp\left[-\left(\frac{t}{\iota^*}\right)^{\beta}\right]
    \label{eq:pow_exp}
\end{equation}
where $f$ is the observable that characterizes the relaxation (e.g. the stress or a structure factor in amorphous solids \cite{Cipelletti2000}), $\iota^*$ a timescale and $\beta$ an exponent corresponding either to a stretched ($\beta<$1) or a compressed ($\beta>$1) exponential. We note that when $\beta$ is significantly lower than 1, a stretched exponential is difficult to distinguish from a logarithmic relaxation.

The $\beta$-value is likely the signature of underlying microscopic processes \cite{Bouchaud2008}. A macroscopic, stretched exponential relaxation may result from the superposition of numerous microscopic exponential relaxations but with a non-trivial distribution of the associated microscopic relaxation times \cite{Johnston2006}. In this case, the macroscopic relaxation is faster than exponential at short timescales, as the result of a profusion of these local relaxations, but much slower ("stretched") at long timescales. Compressed exponential relaxations are less frequent, yet have been observed in soft amorphous media \cite{Cipelletti2000,Cipelletti2005,Bouchaud2008} as well as in glasses below the glass transition temperature \cite{Ruta2012}. 
In contrast to a stretched relaxation, a compressed macroscopic relaxation cannot be explained by the combination of microscopic exponential ones, and rather requires an absence of small microscopic relaxation times in the system \cite{Bouchaud2008}. It is "faster" than exponential in the sense that it occurs abruptly in an avalanche-like manner \cite{Trachenko2021}. 
To explain it, a mechanism involving long-ranged elastic stress redistributions and triggering a succession of concordant rearrangements has been proposed \cite{Cipelletti2000,Cipelletti2005,Bouchaud2008,Trachenko2021}.

\bigskip
As the shear deformation of a granular frictional medium can induce both a degradation of elastic properties, i.e., damage \cite{Karimi2019}, as well as a relaxation of the stress \cite{Hartley2003}, it raises the question of the relation between its elastic and relaxation properties. 
Qualitatively, we could expect damaged systems to relax more easily.

In this paper, we are therefore interested in quantifying these properties for an athermal, dry, frictional granular medium in the critical regime, that is, close to the jamming transition. This transition is of interest in a number of natural contexts (e.g., sea ice, debris flow), for which a continuum representation, based on macroscopic variables and constitutive laws, is still the norm in modeling applications. 
To do so, we represent a dry granular media as an assembly of two-dimensional, bi-disperse circular grains using a molecular dynamics model. This allows (i) tracking the evolution of the elastic modulii at the macroscopic scale of the granular assembly in the course of its imposed shear deformation and therefore define its level of damage, (ii) quantifying its relaxation properties and (iii) establishing possible links between relaxation and elastic properties. 

\bigskip
The paper is structured as follows. First, the simulations are presented along with the model parameters, the boundary and forcing conditions and the different numerical experiment protocols used to estimate the elastic versus the relaxation properties of the simulated granular material. Second, the key results for each type of protocol are presented. Finally, these findings and their consequences are discussed in the context of dry, granular media undergoing large deformations, in the critical regime and in the transition between the dense and critical regimes.

\section{\label{sec:protocol} Simulation Protocols}
In order to track the evolution of the elastic modulii and of the relaxation behavior of a granular media during its deformation, we developed different numerical experiments using the molecular dynamic software LAMMPS \cite{LAMMPS}. The simulated system represents a two dimensional idealized dry granular medium. The experiments can be split into two steps: (i) the preparation, during which the grains are initialized as a granular gas, packed to a packing fraction $\phi_{ini}$, relaxed at a constant volume, and then compressed bi-axially to an initial pressure $P$, and (ii) the simple shear experiment, which is interrupted at different cumulated shear strain values.
From the state of the granular assembly at these interruptions, two types of tests are performed: 
\begin{itemize}
\item small strain, oscillatory tests, in both pure shear and bi-axial compression, which allow measuring its macroscopic elastic properties,
\item relaxation tests under fixed shear strain and volume, which allows estimating its relaxation properties (relaxation time and residual stresses).
\end{itemize}
FIG. \ref{fig:num_protocole} below illustrates these two tests. \newline

\begin{figure}[h]
    \centering
    \includegraphics[width=80mm]{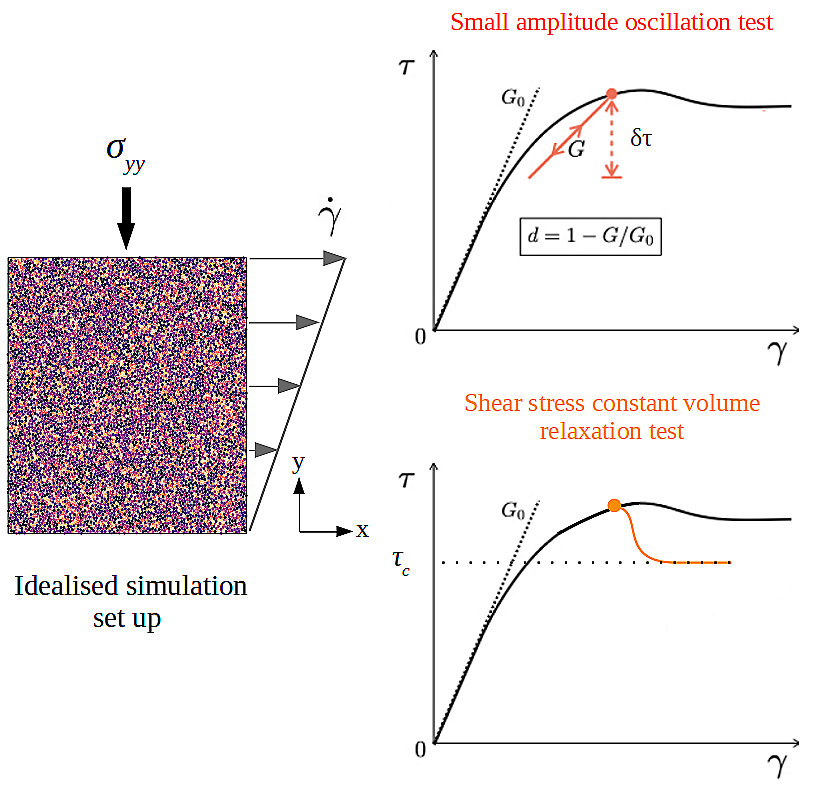}
    \caption{Schematic representations of (left) the simple shear numerical experiment and (right) the tests for the determination of the shear modulus (upper right) and residual shear stress and shear stress relaxation time (lower right). Similar tests are applied to determine the bulk elastic and relaxation properties.}
    \label{fig:num_protocole}
\end{figure}

\subsection{\label{sec:parameters} Simulation parameters}
LAMMPS allows simulating the deformation of a large assembly of particles, or grains, by solving Newton's equations of motion for each grain. Here, each simulated grain is a circular disk of diameter $d$ and density $\rho$, as well as an Hetzian \cite{Mindlin1949} elastic element (i.e., a spring) of normal and tangential stiffness $K_N$ and $K_T$ respectively and characterised by a grain-grain friction coefficient, $\mu_g$. The resonance frequency of the grains defines the normal and tangential oscillation timescales as $t_{N,T}=2 \pi/\omega_{N,T}$ = $2\pi \cdot\sqrt{m_g/\bar{d} \cdot K_{N,T}}$, where $m_g$ is the mass of the grain and $\bar{d}$, the average grain diameter. To avoid numerical issues in the simulation due to strong overlaps and dynamical effects, a microscopic damping term, $\Gamma_{N,T}$, is added to the equation of motion of each grain. $\Gamma_N$ limits the repulsion force in the normal direction while $\Gamma_T$ plays the same role in the tangential direction. 
Elastic forces are ruled by normal and tangential stiffness such that $K_N/K_T = 1.2$, with non-linear Hertzian contacts (Hertz-Mindlin forces) \cite{Zhang2005,Mindlin1949} giving two non-linear repulsive forces between grains, $\vec{F_n}$ and $\vec{F_t}$, that are proportional to the surface of the grains overlap. The elastic modulii of the bulk material that constitutes the grains and the elastic stiffnesses of those grains are linked through $K_N$ = $4G_{2D}/(1-\nu_{2D})$ and $K_T$ = $4G_{2D}/(2 - \nu_{2D})$, where $\nu_{2D}$ and $G_{2D}$ are respectively the 2D Poisson's ratio and shear modulus of the studied material \cite{Zhang2005}.  In molecular dynamics simulations, the discretisation timescale $\Delta t$ should be significantly smaller than the vibrational time of a grain in order to resolve the motion of each grain. Consequently, and following Karimi et al.\cite{Karimi2019}, we set $\Delta t \leq 0.05 \cdot \omega^{-1}_N$ (TABLE \ref{tab:simu_para}).

This work was initially motivated by the mechanical modeling of a particular granular material: sea ice. On geophysical scales, fragmented sea ice can be considered as a two-dimensional, dry granular assembly of ice plates (here, the grains) because of the large aspect ratio between their horizontal extent and thickness, which also implies that hydraulic forces are negligible with respect to friction and collision forces. In the present simulations, we therefore attribute the mechanical properties of our simulated granular medium to the measured properties of sea ice. The Young modulus, $E$, Poisson's coefficient, $\nu$, and density, $\rho$, of this material have been estimated from in situ acoustic wave propagation measurements \cite{Serripierri2022}, from which the shear, $G$, as well as the bulk, $K$, modulii can be easily calculated assuming elastic isotropy in the 2D plane (see TABLE \ref{tab:simu_para}). From these mechanical parameters values, we deduce the normal $K_{N}$ and tangential $K_{T}$ stiffnesses.
From those, we then calculate the vibrational timescales for both the normal and tangential components. Coulombic friction between the grains is ruled by a friction coefficient $\mu_g = 0.7$, that is common to many geomaterials, including sea ice \cite{Weiss2007}. We fix the damping coefficients $\Gamma_N$ and $\Gamma_T$ to $\Gamma_{N,T}= 2 \cdot \omega_{N,T} /\bar{d}$, similar to \cite{Karimi2019}. It is important to note that, although we have set the mechanical properties of our simulated granular medium to be consistent with that of sea ice, the purpose of the present work is wider than the study of this particular material and concerns (2D) dry, frictional granular media in general.

Our simulations represent a dense assembly of $N$ grains, with $N_{max}$ grains of diameter ${d}_{max}$ and $N_{min}$ grains of diameter ${d}_{min}$. The proportion of large versus small grains is set to $N_{max} / N_{min}$ = 1 and the grain diameter ratio is set to ${d}_{max} / {d}_{min}$ = 1.4. We recall that in our simulations, the material is considered as purely 2-dimensional. Hence there is no motion in the out-of-plane dimension and the grains are considered as being disks of zero thickness. The simulation box geometry is set as "triclinic", allowing deformation along the $O_{xx}$, $O_{yy}$ and $O_{xy}$ directions. Bi-periodic boundary conditions are set along the $O_{xx}$ and $O_{yy}$ directions, allowing grains to "flow" through the boundary of the domain and re-enter the box through the opposite side. The type of periodic boundary condition is chosen such that to minimize boundary effects. It consists in adding a position proportional to the displacement of the lower and upper boundary to any grain re-entering the domain \cite{LAMMPS}.

In what follows, all the parameters and variables have been made adimensional with respect to the average grain diameter, $\bar{d}$, the Young modulus, $E$, of the simulated material, the resonance time of a grain (in the normal direction), $t_N$, and the density of the material, $\rho$. Adimensional parameters and variables are denoted with the notation $\tilde{}$ and the adimensional formulas and values used are summarized in TABLE \ref{tab:simu_para}.

\renewcommand{\arraystretch}{1.5}
\begin{table*}[t!]
\centering
\begin{tabular}{|c c c c|} 
 \hline
 Parameter & Notation& Formula & Value \\
 \hline\hline
 Young modulus of a grain& $\tilde{E}_g$& $E_g/E_g$& $1$\\
 Poisson coefficient of a grain& $\nu_g$& & $0.3$\\
 Normal oscillation time of a grain& $\tilde{t}_N$ & $t_N / t_N$ & $1$\\
 Tangential oscillation time of a grain& $\tilde{t}_T$& $t_T/t_N$&$1.1$\\
 Average grain diameter & $\tilde{d}_g$& $\bar{d}_g/\bar{d}_g$& $1$\\
 Grain Density & $\tilde{\rho}_g $& $\rho_g / \rho_g$& $1$ \\ 
 Average grain mass & $\tilde{m}_g$ & $\bar{m}_g \cdot \bar{d}^2/\rho_g $ & $7.85 \cdot 10^{-1}$\\
 $y-$ dimension of the system & $\tilde{l}_y$ & $l_y/ \bar{d}$ & $1 \cdot 10^2$ \\
 Normal stiffness & $\tilde{K}_N$ &  $K_N / E_g$ & $2.2$ \\
 Tangential stiffness & $\tilde{K}_T$ & $K_T / E_g$ & $1.81$ \\
 Normal damping coefficient & $\tilde{\Gamma}_T$ & $\Gamma_N \cdot t_N \cdot \bar{d}$ & $9.98 \cdot 10^{1}$\\
 Tangential damping coefficient & $\tilde{\Gamma}_T$ & $\Gamma_T \cdot t_N \cdot \bar{d}$ & $9.42 \cdot 10^{1}$\\
 Grain to grain friction & $\mu_g$& & $0.7$\\
 Simulation time step & $\Delta \tilde{t}$ & $\Delta t/ t_N$ & $1.4 \cdot 10^{-3}$\\
 Simulation time step for small strain oscillatory tests & $\tilde{\Delta t}_{p}$& $\Delta t_{p} / t_N$& $1.4 \cdot 10^{-5}$ \\
 Time of loading of small strain oscillatory tests & $\tilde{T}_p$ & $T_p / t_N$ & $7.1 \cdot 10^{-4}$ \\
 Duration of relaxation tests & $\tilde{T}_{r}$& $t_{r} / t_N$& $5.7 \cdot 10^{3}$ \\
 Microscopic relaxation time for the grains & $\tilde{\lambda}$ & $\lambda / t_N$ & $1 \cdot 10^{-2}$ \\
 \hline\end{tabular}
\caption{Adimensional simulation parameters for the simple shear experiment,small strain oscillatory tests and relaxation tests.}
\label{tab:simu_para}
\end{table*}

\subsection{\label{sec:prep_protocol} Preparation protocol}
The objective of the preparation stage is to obtain a dense, yet stable assembly of grains. To initialize this system, a dilute granular gas with random positions for $N$ grains is created. The initial packing fraction is chosen low ($\phi_{ini} = 0.23$) in order to minimize the probability of grain overlaps. Next, a compaction step is performed by applying a bi-axial compression up to a targeted, high packing fraction of $\phi = 0.8$, with the grain-grain friction coefficient ($\mu_g$) and the tangential contact force ($\tilde{K}_T$) switched off (i.e., set to 0). A full relaxation of the system at constant volume over a duration $\tilde{t}_{relax}$ = $7 \cdot 10^4$, is then carried to minimize grain overlaps, dissipate internal stress and ensure the simulation stability in the next steps. Finally, the system is bi-axially compressed to a given pressure, $\tilde{P}$ = $0.5 \cdot (\tilde{\sigma}_{xx} + \tilde{\sigma}_{yy})$. Two different values are explored: $1.014 \cdot 10^{-4}$ and $1.8 \cdot 10^{-4}$, referred to in the following as the applied normal stress $\tilde{\sigma}_{yy}$. During this sample preparation step, the simulation time step ($\Delta \tilde{t}$) is increased by a factor $10$ in order to reduce computation times. All of the simulation results presented in this paper are the averages of an ensemble of 30 simulations on systems initialized with different random positions of the grains.

\subsection{\label{sec:shear_test} Simple shear experiment}
The preparation step completed and the initial value of pressure reached, a simple shear loading is prescribed on the granular assembly while applying a constant normal stress ($\tilde{\sigma}_{yy}$ = $1.014 \cdot 10^{-4}$ or $1.8 \cdot 10^{-4}$) on the upper and lower boundaries of the simulation domain. A constant strain rate $\tilde{\dot{\gamma}}$ is applied until the system reaches a final shear strain of $0.3$. Practically, this strain rate is imposed by increasing the tilt of the box. Five values of the shear strain rate within the range $\tilde{\dot{\gamma}} \in [2.625 \cdot 10^{-7}$; $2.625 \cdot 10^{-5}]$ have been tested. Simulations using the same values of the inertial number $I$ obtained from different couples of values of $\tilde{\dot{\gamma}}$ and $\tilde{\sigma}_{yy}$ have also been performed (see TABLE \ref{tab:load_cond}) in order to determine to what extent $I$ can describe the behavior of the system.
During the simple shear experiment, both the grain-grain friction ($\mu_g$) and the tangential contact force $\tilde{K}_T$ are switched on. The ratio of normal grain stiffness $\tilde{K}_N$ and applied normal stress $\tilde{\sigma}_{yy}$ is set $> 10^4$ in order to remain above the rigid grain limit \cite{daCruz2005}.

For all simulations, the experiment is interrupted at $19$ chosen values of shear strain and the state of the granular assembly (grain positions, overlaps, diameters, etc.) and the shape of the domain are saved. The small strain oscillatory tests and relaxation tests are performed using these saved macroscopic and microscopic configurations as initial conditions (see FIG. \ref{fig:num_protocole}).

\renewcommand{\arraystretch}{1.5}
\begin{table}[b!]
\centering
\begin{tabular}{|c c c c c|}
 \hline
 $I$  & & $\tilde{\dot{\gamma}}$ $(\tilde{\sigma}_{yy}) = 1 \cdot 10^{-4})$ & & $\tilde{\dot{\gamma}}$ $(\tilde{\sigma}_{yy} = 1.8 \cdot 10^{-4})$\\
 \hline\hline
 $5 \cdot 10^{-3}$ & & $2.625 \cdot 10^{-5}$ & & $3.5 \cdot 10^{-5}$\\ 
 $1 \cdot 10^{-3}$ & & $5.25 \cdot 10^{-6}$ & & $7 \cdot 10^{-6}$ \\
 \hline
 $5 \cdot 10^{-4}$ & & $2.625 \cdot 10^{-7}$ & & $3.5 \cdot 10^{-6}$ \\ 
 $1 \cdot 10^{-4}$ & & $5.25 \cdot 10^{-8}$ & & $7 \cdot 10^{-7}$\\
 $5 \cdot 10^{-5}$ & & $2.625 \cdot 10^{-8}$ & & $3.5 \cdot 10^{-7}$ \\ 
 \hline
\end{tabular}
\caption{Imposed loading conditions in the simple shear numerical experiments, for each explored value of the inertial number $I$.}
\label{tab:load_cond}
\end{table}

\subsection{\label{sec:oscillatory_tests} Small strain oscillatory tests}
These tests consist in imposing either a pure shear deformation or a bi-axial compression, oscillating around the initial configuration following a sine function of very small amplitude $\gamma_A = 2 \cdot 10^{-7}$ and $\epsilon_{kk,A} = 2 \cdot 10^{-7}$ respectively. The oscillation period is set to $\tilde{T}_p$ (see TABLE \ref{tab:simu_para}) and the number of loading/unloading cycles to $N_{cycle} = 40$. The simulation time step during these tests, $\tilde{\Delta t}_p = 1.4 \cdot 10^{-5}$, is set to a smaller value than for the shear experiment, in order to resolve the effect of these oscillations on the granular assembly. To avoid inertial effects inherited from the imposed macroscopic shear deformation, the initial velocity of the grains is set to $\tilde{v}(t=0)$ = $0$ at the beginning of each oscillatory test. Tests performed for different values of $I$ showed that this lowers the uncertainty associated with the measurement of the elastic modulii and reduces the observed hysteresis. As discussed in section \ref{sec:result}, these tests allow probing the elastic properties of the granular configuration, i.e. estimating $\tilde{G}$ and $\tilde{K}$ \cite{Digby1981,Gammon1983,Karimi2019}.

\subsection{\label{sec:relaxation_tests} Relaxation tests}
Relaxation tests are performed under constant volume. From the saved initial micro-macro configurations, the system is allowed to relax with bi-periodic boundary conditions. This approach allows fixing the shape of the box without any remapping of the grains near the boundaries, thereby avoiding artificial changes in the topological configuration of the system. A characteristic time of relaxation $ \tilde{t}^*$ can be computed from the relaxation tests, following Eq. \ref{eq:mean_relax_t}, as further detailed in section \ref{subsec:relax}. The duration of the relaxation tests is $\tilde{T}_{r}$, such that $\tilde{t}^* << \tilde{T}_r$ (see TABLE \ref{tab:simu_para}), which ensures that the system reaches a macroscopic steady state.

\subsection{\label{sec:timescales} Timescales}
\begin{figure}[h]
    \centering
    \includegraphics[width=80mm]{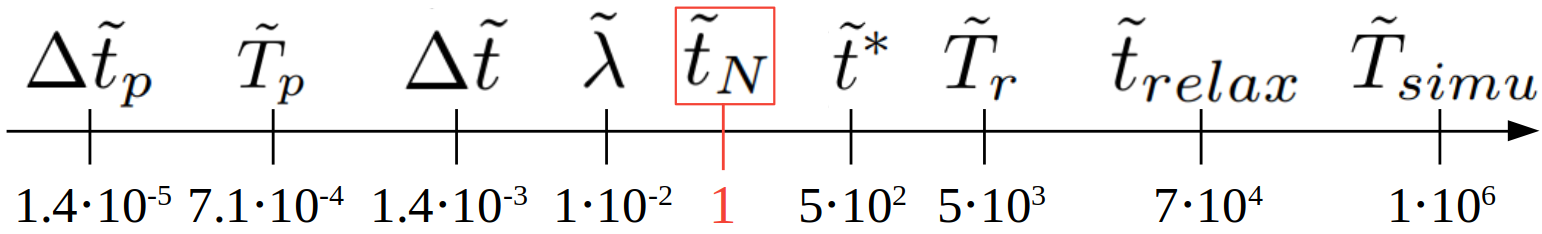}
    \caption{Relevant timescales in our simulations.}
    \label{fig:t_simu}
\end{figure}

At this point, a comparison of the main timescales involved in our simulations appears useful. As explained above, the grain mass and grain contact stiffnesses $\tilde{K}_{N,T}$ set the corresponding vibrational timescales $\tilde{t}_{N,T}$. We choose $t_{N}$ as our reference timescale. Hence $\tilde{t}_{N}=1$, by construction, while $\tilde{t}_{T}$ = $\frac{t_T}{t_N}$ is slightly larger (see TABLE \ref{tab:simu_para}). Three other timescales emerge from the protocol described above: 1) the oscillation period $\tilde{T}_p$ of the oscillatory tests, 2) the grain scale dissipation time $\tilde{\lambda}$ and 3) the characteristic relaxation timescale of the whole system, $\tilde{t}^*$, estimated from the relaxation tests. All of these timescales should be greater than the simulation time step $\Delta \tilde{t}$ in order to ensure the correct resolution and monitoring of the physical processes of interest. The grain scale dissipation time is estimated from the microscopic input damping parameters $\tilde{\Gamma}_{N,T}$ as $\tilde{\lambda} \sim 1 / (\Gamma_N \cdot \bar{d}) \cdot (1/t_N)$. Therefore, $\tilde{\lambda}$ is set such as $\tilde{\lambda} < \tilde{t}_N$, as the micro-scale relaxation time allow the partial viscous relaxation of stresses associated with grain overlaps while ensuring stability of the simulations. The time of loading of the small strain oscillatory test is set such that $\tilde{t}_N > \tilde{\lambda} \gg \tilde{T}_{p}$. This avoids soliciting the grains at their resonance time ($\tilde{t}_N$), and avoids measuring the grains micro-damping associated to $\tilde{\lambda}$. It also allows defining two discretisation time steps, one for the simple shear and relaxation tests, such that $\Delta \tilde{t} \ll \tilde{\lambda}$, another one for the small strain oscillatory tests, such that $\Delta \tilde{t}_{p} \ll \tilde{T}_p$ (see FIG. \ref{fig:t_simu} and TABLE \ref{tab:simu_para}).

We also verified that the measured macroscopic relaxation timescales of the whole assembly of grains, $\tilde{t}^*$, are significantly larger than the other timescales, but significantly smaller than the duration of the relaxation test $\tilde{T}_r$. We therefore have $\Delta \tilde{t}_p < \tilde{T}_p < \Delta \tilde{t} < \tilde{\lambda} < \tilde{t}_N < \tilde{t}^*$ which is summarized in FIG. \ref{fig:t_simu}.

\section{\label{sec:result} Results}

\subsection{\label{subsec:size_eff} Size effects}

\begin{figure*}[!t]
    \begin{subfigure}{0.99\textwidth}
        \centering
        \includegraphics[width=\textwidth]{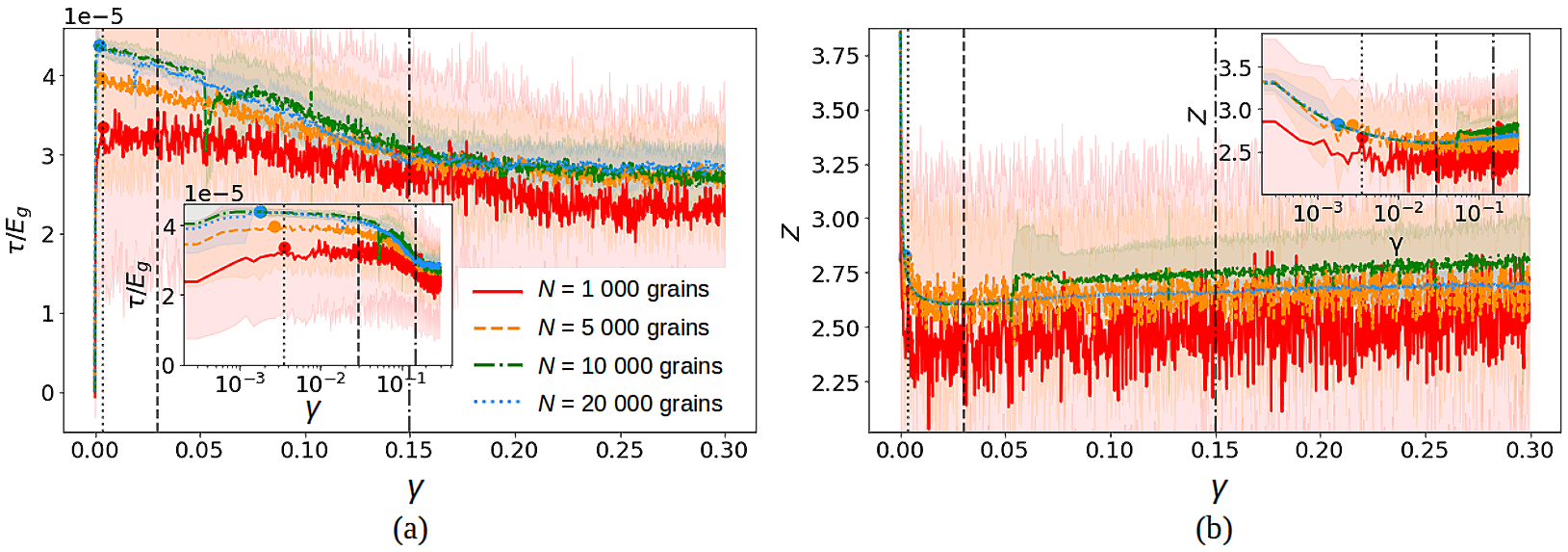}
    \end{subfigure}
    \caption{Evolution of (a) the shear stress, $\tilde{\tau}$, and (b) the coordination number, $Z$, as a function of the imposed shear strain $\gamma$, for different system sizes ($N = 1000$, $5000$, $10000$ and $20000$) in the critical regime ($I \approx 1 \cdot 10^{-4}$, $\tilde{\dot{\gamma}} = 5.25 \cdot 10^{-7}$, $\tilde{\sigma}_{yy} = 1 \cdot 10^{-4}$). Coloured dots indicate the peak stress $\tilde{\tau}_{max}$ in each simulation and the associated coordination number. The insets show the same data using a semi-log scale, to focus on the initial behavior. The vertical lines separate the different stages of deformation: the initial elastic (dot lines), the meta-stable (dashed lines), strain-softening (long dash-dot) and saturation stage.}
    \label{fig:sample_size_cs}
\end{figure*}

\begin{figure}[!h]
    \centering
    \begin{subfigure}{0.49\textwidth}
        \includegraphics[width=\textwidth]{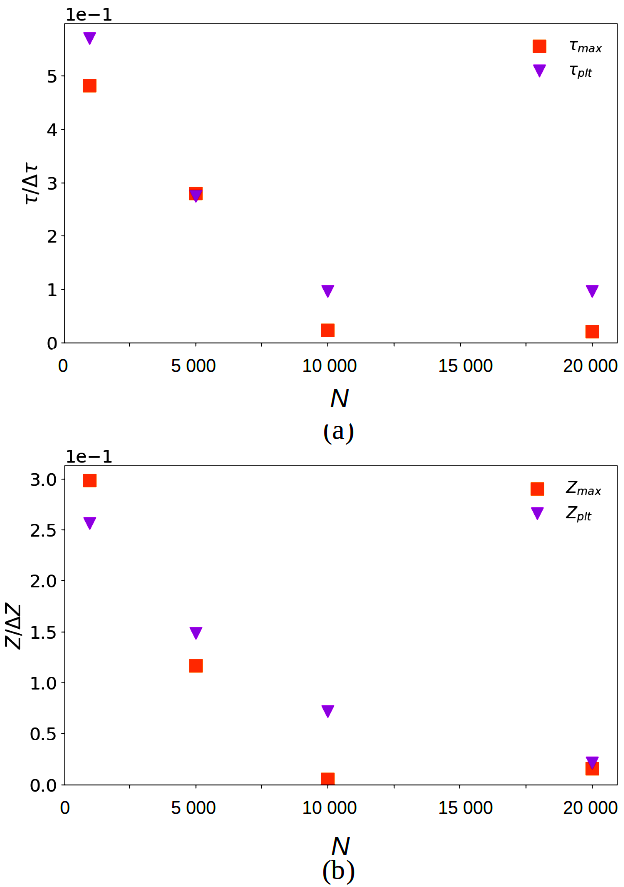}
    \end{subfigure}
    \caption{Variability among 30 simulations with different initial grain positions of (a) the value of peak shear stress, $\tilde{\tau}_{max}$, and of the shear stress measured at $\gamma = 0.3$, $\tilde{\tau}_{plt}$ and (b) corresponding variability of the coordination numbers $Z_{max}$ and $Z_{plt}$ 
    for system sizes of $N$ = $1000$, $5000$, $10000$ and $20000$ grains.
    }
    \label{fig:size_eff_err_cs}
\end{figure}
Both laboratory experiments and numerical simulations of granular media show that the microscopic observables and the macroscopic mechanical parameters are subjected to size effects \cite{Dufour2007,TMiller2013,Weiss2014,Cil2016}. To account for and characterize these effects, we first compared simple shear numerical experiments in the critical regime ($I = 1 \cdot 10^{-4}$) performed on systems involving a different number of grains ($N = 1000$, $5000$, $10000$ and $20000$). On FIG. \ref{fig:sample_size_cs}, we represent the shear stress as well as the evolution of the average coordination number $Z$ as a function of the shear strain for these different system sizes. We recall that these results have been obtained by averaging 30 simulations initialized with different grain positions.

The evolution of the macroscopic shear stress $\tilde{\tau}$ follows a similar behavior for all sample sizes $N$. It is characterized by an initial increase upon loading, up to a peak stress $\tilde{\tau}_{max}$ observed around $\gamma \approx 1\cdot 10^{-3}$, a strain softening ($\gamma \approx 0.02 \to 0.15$) and a stationary regime at high deformation ($\gamma > 0.15$), for which a constant value of shear stress is reached (see FIG. \ref{fig:sample_size_cs}a). The slight gaps in stress observed for $N =10000$ and $20000$ grains at $\gamma = 2.5 \cdot 10^{-2}$ and $5 \cdot 10^{-2}$ are due to simulation restarts. The peak stress $\tilde{\tau}_{max}$ slightly increases with the system size $N$, but the difference between $N=10000$ and $N=20000$ is minimal. An even smaller size effect is observed for the shear stress reached in the stationary regime. 

The coordination number quickly evolves from $Z \approx 3.8$ at the onset of loading to a residual value of $Z \approx 2.7$ for $\gamma > 0.02$ (i.e. beyond the peak stress), whatever the system size (see FIG. \ref{fig:sample_size_cs}b). Differences between the system sizes $N=10000$ and $N=20000$ are hardly discernible. 

To explore further these size effects, we look for different system sizes at the evolution of the variability, defined as the standard deviation over the 30 realizations with different initial grain positions of $\tilde{\tau}$ and $Z$ at the peak stress (subscript ''max'') and at the end of the shear deformation experiment (subscript ''plt'', see FIG. \ref{fig:size_eff_err_cs}a and \ref{fig:size_eff_err_cs}b). As expected \cite{Weiss2014}, these variabilities both decrease with increasing system size. For $N \geq 10,000$, they are below 10$\%$.

In conclusion, size effects appear limited for $N \geq 10000$. Therefore, in order to minimize both size effects and computational times, we used assemblies of $N=10000$ grains in all of the following simulations.

\subsection{\label{muI} Comparison with $\mu(I)$}

\begin{figure}[t]
    \centering
    \includegraphics[width=85mm]{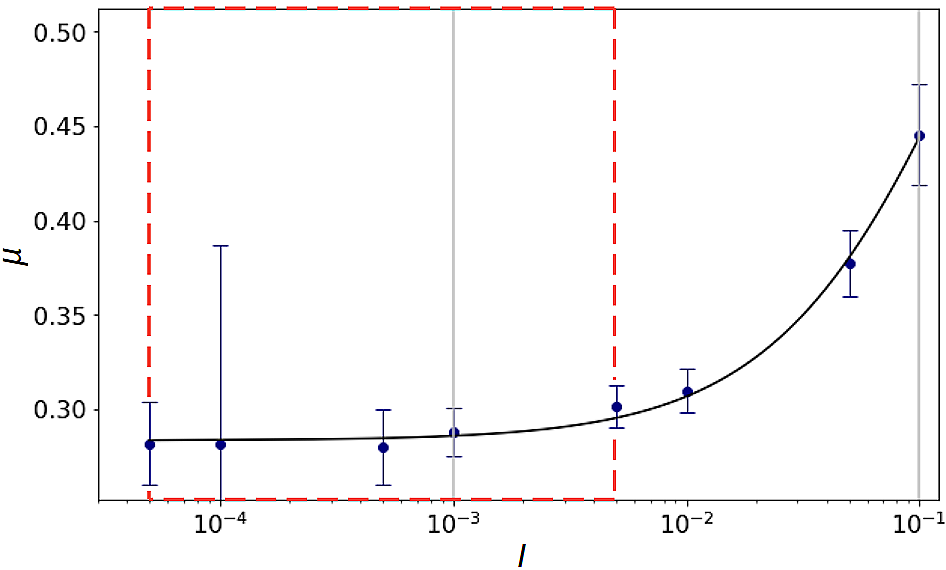}
    \caption{Evolution of the macroscopic effective friction coefficient $\mu$ as a function of the inertial number, $I$ for systems with $N = 10000$ grains. The red box indicates the range of $I$ values investigated in the remaining simulations. The black solid line shows a fit to a $\mu(I)$ rheology that uses $\mu_1 = 0.28 \pm  2.5 \cdot 10^{-6}$, $\mu_2 = 0.75 \pm 1.0 \cdot 10^{-2}$ and $I_0 = 0.19 \pm 3.6 \cdot 10^{-3}$.}
    \label{fig:mu_I_results}
\end{figure}

To validate our representation of a granular media as a 2D assembly of circular particles with non-linear elastic contacts and inter-particle friction, we compared our simulation results to the prediction of the macroscopic effective friction coefficient $\mu$ of the $\mu(I)$ framework, which is known to be appropriate in the case of idealized, dry frictional granular media \cite{GDRMidi2004}. To do so, we performed simulations with $N = 10000$ grains strained under a constant normal stress of $\tilde{\sigma}_{yy} = 1 \cdot 10^{-4}$ but various imposed shear strain rate $\tilde{\dot{\gamma}}$ (see TABLE \ref{tab:load_cond}) and computed the value of  $\mu = \tilde{\tau} / \tilde{\sigma}_{yy}$ as a function of the inertial numbers $I$, ranging from the critical to the dense flow regime. As shown in FIG. \ref{fig:mu_I_results}, our results are in good agreement with a $\mu(I)$ prediction with parameters $\mu_1 = 0.28 \pm  2.5 \cdot 10^{-6}$ and $I_0 = 0.19 \pm 3.6 \cdot 10^{-3}$ in the dense flow and critical regimes. Although these parameters are material- and geometry-dependent, these values fall within the ranges reported in previous studies. For instance, typical values of $\mu_1 = 0.38$ and $I_0 = 0.3$ have been reported for 3D glass beads assemblies \cite{Pouliquen2006} and values of $\mu_1 = 0.13$ and $I_0 = 6.8 \cdot 10^{-3}$ have been obtained with a very similar molecular dynamics model representing 2D circular disks with a large polydispersity \cite{Herman2022}. Also as expected, the value of $\mu_1$, which characterizes the effective friction of the whole grain assembly in the critical regime, is significantly lower than the inter-particle friction coefficient $\mu_g=0.7$.
    
In this work, inertial numbers in the range $[5 \cdot 10^{-5};5 \cdot 10^{-3}]$ are explored to allow a better understanding of the granular behavior in the critical regime and at the transition between the critical and the dense flow regimes (hereinafter simply referred to as the dense flow regime).

\subsection{\label{subsec:sh} Simple shear experiments}

\begin{figure}[!h]
    \begin{subfigure}{0.48\textwidth}
                \includegraphics[width=\textwidth]{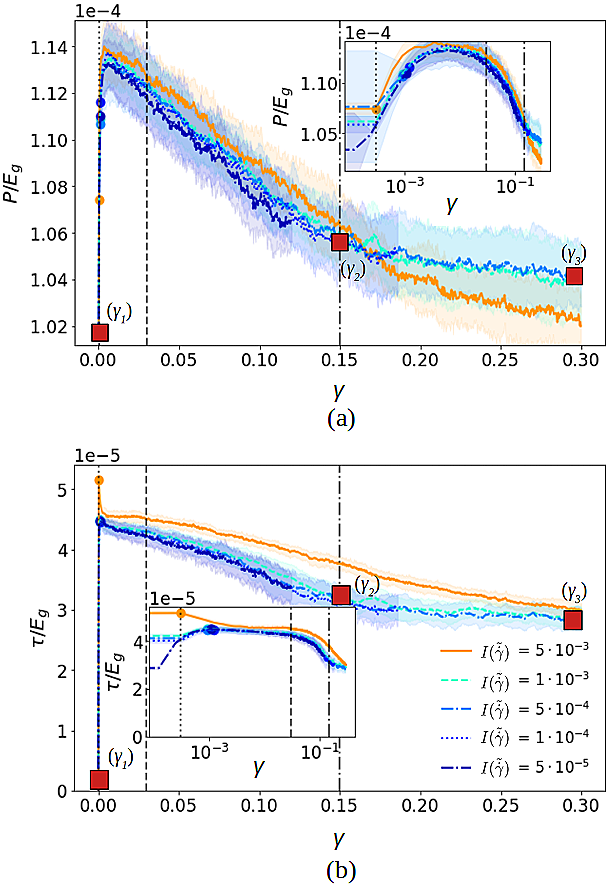}
    \end{subfigure}
    \hfill
    \caption{Evolution of (a) the macroscopic pressure $\tilde{P}$ and (b) the shear stress $\tilde{\tau}$ as a function of the shear strain $\gamma$ during the simple shear test, using $N = 10000$, $\tilde{\sigma}_{yy}= 1.0 \cdot 10^{-4}$ and four different values of the prescribed shear strain rate $\tilde{\dot{\gamma}}$. Coloured dots on both panels indicate the shear stress peaks. Red squares labelled ($\gamma_1$), ($\gamma_2$), and ($\gamma_3$) refer to the levels of deformation at which the micro-mechanics is represented in FIG. \ref{fig:simple_shear_mm}, with $\gamma_1 = 0$, $\gamma_2 = 0.15$ and $\gamma_3 = 0.3$.}.
    \label{fig:simple_shear_stress}
\end{figure}

\begin{figure}[!h]
    \begin{subfigure}{0.5\textwidth}
        \includegraphics[width=\textwidth, right]{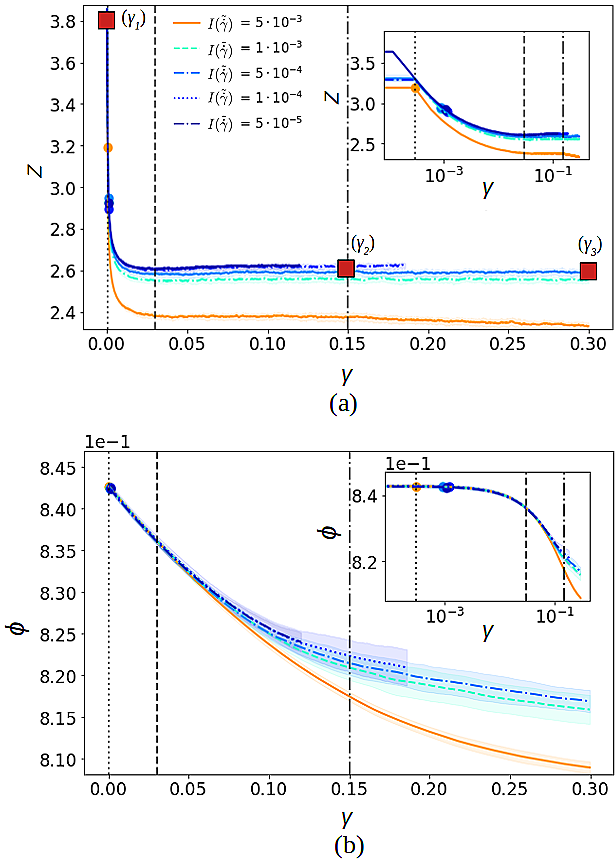}
    \end{subfigure}
    \hfill
    \caption{Evolution of (a) the coordination number $Z$ and (b) the packing fraction $\phi$ as a function of the shear strain $\gamma$ during the simple shear experiment, using $N = 10000$, $\tilde{\sigma}_{yy} = 1.0 \cdot 10^{-4}$ and four different values of the prescribed shear strain rate $\tilde{\dot{\gamma}}$. Coloured dots on both panels indicate the value of $Z$ and $\phi$ at the peak stress. The red squares with labels ($\gamma_1$), ($\gamma_2$) and ($\gamma_3$) refer to the the levels of deformation at which the micro-mechanics is analysed.}
    \label{fig:simple_shear_vol}
\end{figure}

\begin{figure}[!t]
    \begin{subfigure}{0.49\textwidth}
        \includegraphics[width=\textwidth]{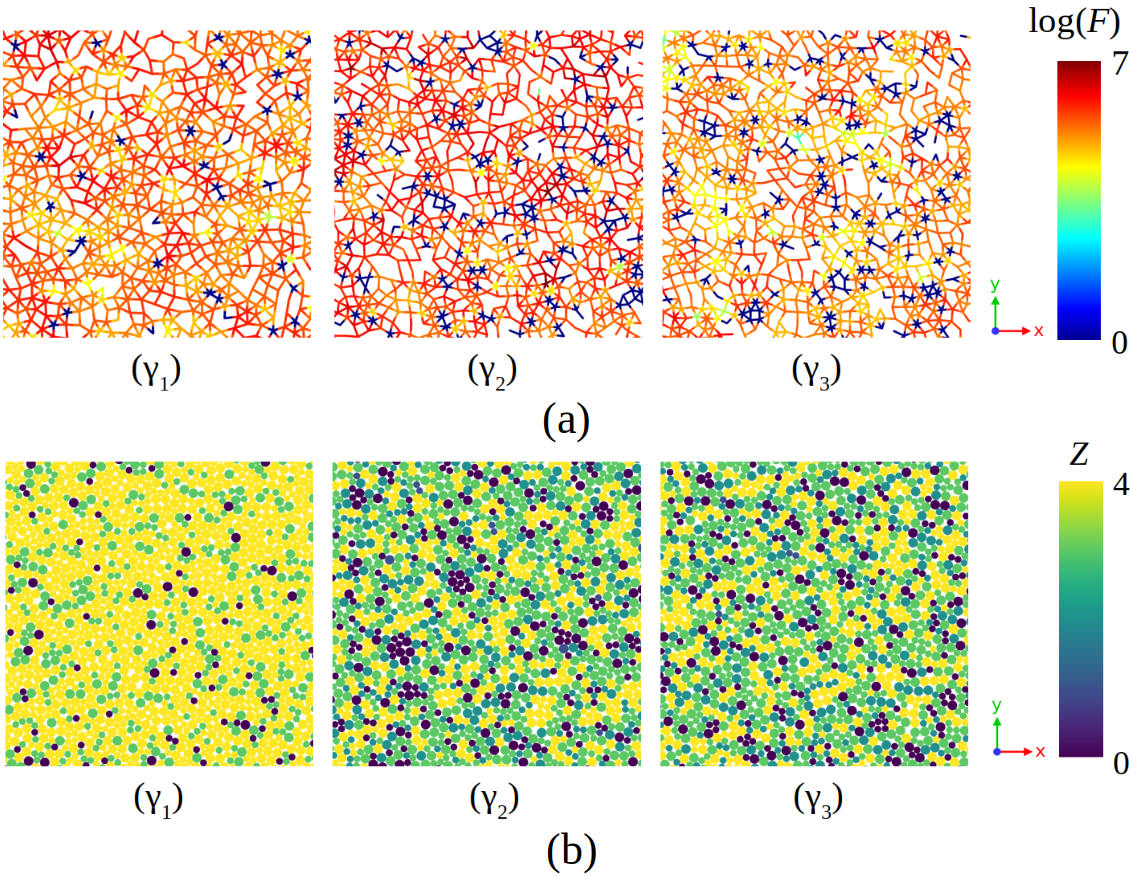}
    \end{subfigure}
    \caption{Micro-mechanical evolution of (a) the force chains and (b) the coordination number during a simulation with $N = 10 000$ and $I = 5 \cdot 10^{-4}$ at different levels of deformation: $\gamma_1$ = $0$, $\gamma_2$ = $0.15$ and $\gamma_3$ = $0.3$.}
    \label{fig:simple_shear_mm}
\end{figure}

In this section, we discuss in more details the macroscopic behavior of our simulated granular medium for different values of the inertial number falling within the critical and dense flow regimes in terms of the imposed macroscopic shear stress $\tilde{\tau}$, pressure $\tilde{P}$, average coordination number $Z$ and packing fraction $\phi$ (see FIG. \ref{fig:simple_shear_stress} and \ref{fig:simple_shear_vol}). As these macroscopic (i.e., averaged) variables do not bring direct information on the microscale evolution of the system, we complete our analysis with a monitoring of the magnitude of the force chains and local coordination numbers throughout the experiments (see FIG. \ref{fig:simple_shear_mm}). Many authors performed detailed statistical analyses of the micro-mechanical behavior of frictional granular media in relation to their macroscopic behavior (e.g. \cite{Roux2002,Radjai2002,Hartley2003}). This is not our main objective here. Instead, we characterise this evolution in a more qualitative manner.
To improve readability, we zoom over a region of about $1000$ grains at the centre of the $N = 10000$ system. It should be emphasized that the relative spatial homogeneity of force chains and coordination numbers which we observe within our granular assemblies (see e.g., FIG. \ref{fig:simple_shear_mm}) validates the use of our $N = 10000$ system size as a representative "volume" to explore the evolution of macroscopic elastic and relaxation properties during its deformation.
   
FIG. \ref{fig:simple_shear_stress}a and \ref{fig:simple_shear_stress}b show a similar evolution, already described in section \ref{subsec:size_eff}, for the macroscopic pressure and shear stress. 
It is noteworthy to mention that by construction, the normal stress $\tilde{\sigma}_{yy}$ is maintained constant during the experiment. Therefore, the evolution of the pressure $\tilde{P}$ only results from an evolution of $\tilde{\sigma}_{xx}$. Even in the stationary stage observed at large strains, $\tilde{P}$ remains larger than the applied $\tilde{\sigma}_{yy}$, but this difference is of a few per cents. 

FIG. \ref{fig:simple_shear_stress}b suggests that, in the critical regime ($I \leq 10^{-3}$), the evolution of both the macroscopic shear stress and pressure is $I-$independent over the entire deformation range explored. As the different values of $I$ here are obtained by varying the prescribed strain rate, $\tilde{\dot{\gamma}}$, this implies that the these evolutions are strain-rate independent.  
It is not the case outside of the critical regime. Indeed, comparing the curves for $I \leq 10^{-3}$ and $I=5 \cdot 10^{-3}$, the peak stress appears larger in the dense than in the critical regime and is also followed by a different strain-softening stage.

As already shown in Section \ref{subsec:size_eff}, the mean coordination number $Z$ rapidly drops during the simple shear experiment, from $Z \simeq 3.8$ to a plateau at $\gamma>0.02$. 
FIG. \ref{fig:simple_shear_mm}b and \ref{fig:simple_shear_mm}a track this evolution at the micro-scale, in the case of $I= 5 \cdot 10^{-4}$ (critical regime). 
At the end of the preparation step, most of the grains are in contact with 4 neighbors (i.e., $Z = 4$) and form a percolating network in which there is only a few rattlers (grains with no frictional contacts, able to freely move within a short distance of neighboring grains). 
At the end of the strain softening stage ($\gamma=0.15$), this network appears broken. 
Non-rattlers grains organize as isolated clusters and the number of rattlers increases. 
At the end of the experiment ($\gamma=0.3$), which corresponds to the macroscopic stationary stage, the micromechanics still slightly evolves, with topological rearrangements that decrease the magnitude of force chains (see FIG. \ref{fig:simple_shear_mm}a).
Again, this evolution and, in particular, the plateau value of $Z$ evolves only marginally with the applied strain rate (i.e., $I$) in the critical regime, but decreases significantly once entering the dense flow regime, consistent with a less solid-like and more fluid-like behavior (see FIG. \ref{fig:simple_shear_vol}a). 
Similarly, if the residual value of the packing fraction $\phi$ at $\gamma=0.3$ is almost independent of the applied shear rate (i.e., $I$) in the critical regime, it is not when comparing the critical and dense flow regimes (see FIG. \ref{fig:simple_shear_vol}b). 
The much more pronounced decrease of $\phi$ observed in this last regime indeed points to a more important dilation of the system. In all cases however, $\phi$ evolves initially much less rapidly than $Z$. The rapid drop in $Z$ at the onset of shear loading points to a strain-induced "dejamming" of the system. 
To test our hypothesis that this deformation-induced "dejamming" takes place through successive topological rearrangements of the grains, in section \ref{subsec:ela_prop}, we quantify the evolution of the elastic and relaxation properties of our simulated granular media throughout the simple shear experiment.

To summarize, the behavior of our simulated granular medium under simple shear can be divided into four stages, which are delimited by vertical lines on all figures representing the macroscopic evolution as a function of the applied deformation $\gamma$ (e.g., see FIG. \ref{fig:sample_size_cs}a):

\begin{enumerate}
\item A first, elastic stage, corresponding to a rapid increase of the stresses ($\tilde{P}$ and $\tilde{\tau}$), ending with a brutal dejamming as the coordination number drops from $Z \approx 4$ to $Z < 3$. 
It is worth noting that the packing fraction $\phi$ remains relatively unchanged during this initial stage, meaning that the system neither dilates nor contracts. The loss of grain contacts (FIG\ref{fig:simple_shear_vol}a) is due to numerous granular rearrangements allowing the system to unjam. As it will be shown below in section \ref{subsec:ela_prop}, these rearrangements lead also to a significant decrease of the elastic moduli, i.e. a damage of the granular system. Consequently, non-linear elasticity emerges during this early stage of deformation.
\item As the applied shear strain reaches $\gamma \simeq 1 \cdot 10^{-3}$, the shear stress and pressure reach their peak values, and macroscopic plasticity follows. The stress then stabilizes during a "meta-stable" stage. During this deformation stage, granular rearrangements continue while the coordination number finally reaches a plateau around $Z \to 2.5$. The packing fraction $\phi$ decreases but remains strain-rate invariant, whereas the evolution of all the others macroscopic or averaged variables ($\tilde{\tau}$, $\tilde{P}$, $Z$) differs between the dense flow and the critical regime (FIG \ref{fig:simple_shear_stress}, \ref{fig:simple_shear_vol}).
\item This is followed by a strain-softening stage between $\gamma \simeq 0.03$ and $0.15$ (FIG. \ref{fig:simple_shear_stress}). While $Z$ remains constant until the end of the simulation, the packing fraction $\phi$ continues to decrease during this deformation stage, although at a rate that depends on the applied shear rate, and so on the inertial regime.
\item Finally, beyond $\gamma = 0.15$, the granular assembly reaches a stationary stage, during which $\tilde{P}$, $\tilde{\tau}$ and $Z$ remain constant, while the packing fraction continues to slightly decrease. \\
\end{enumerate}

\begin{figure}[!h] 
    \begin{subfigure}{0.49\textwidth}
        \includegraphics[width=\textwidth]{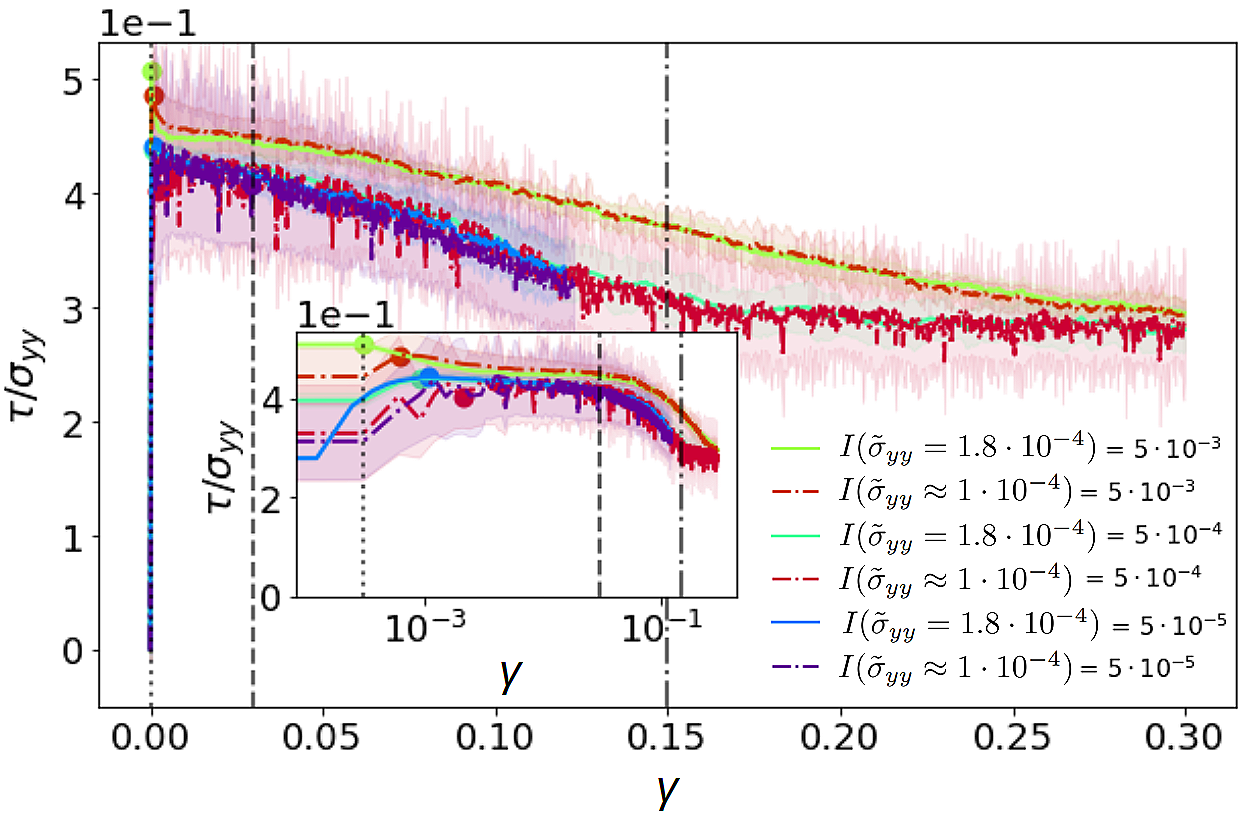}
    \end{subfigure}
    \caption{Evolution of effective friction $\mu = \tau / \sigma_{yy}$ as a function of the shear strain $\gamma$ during the simple shear experiment using $N = 10000$, normalized by the applied normal stress $\tilde{\sigma}_{yy}$,  and three different values of $I$ obtained for different couples of values of $\tilde{\sigma}_{yy}$ and $\tilde{\dot{\gamma}}$. Coloured dots on both panels indicate the peak shear stress.}
    \label{fig:simple_shear_stress_press}
\end{figure}
By construction, the $\mu(I)$ rheology \cite{Jop2006,Pouliquen2006} characterizes the shear resistance of the granular medium in the fourth, or stationary, stage. In other words, it cannot describe the transient stages (1 to 3). In addition, the prediction of $\mu(I)$ in the critical state is that of a $I$-independent macroscopic friction, $\mu=\mu_1$, in agreement with our results (FIG. \ref{fig:mu_I_results}). This raises the interesting question of the possible relevance (or not) of the inertial number $I$ in controlling the mechanical behavior of our dry, frictional granular assemblies during the transient stages.

\begin{figure}[h]
    \begin{subfigure}[!h]{0.5\textwidth}
        \includegraphics[width=\textwidth, right]{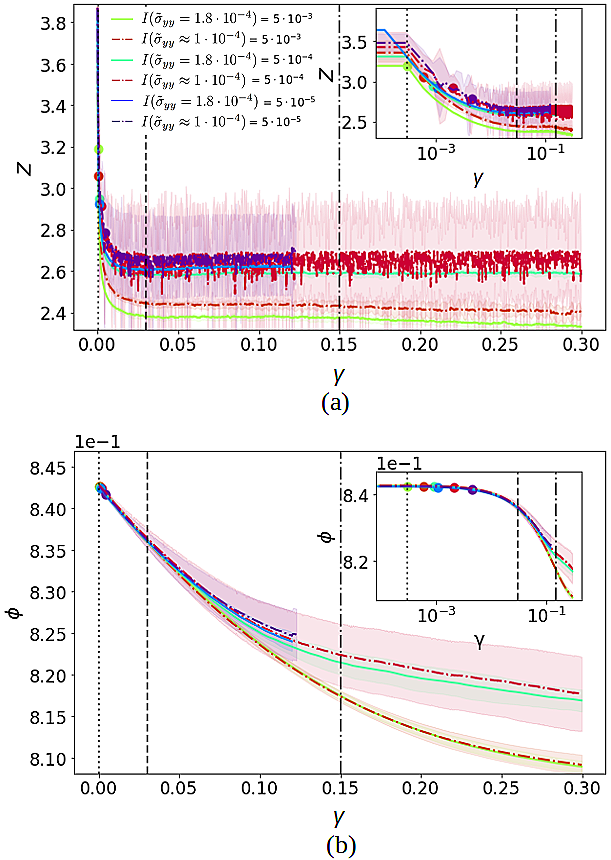}
    \end{subfigure}
    \caption{Evolution of (a) the coordination number $Z$ and (b) the packing fraction $\phi$ as function of the shear strain $\gamma$ during the simple shear experiment using $N = 10000$ and $I = 5 \cdot 10^{-3}$, $I = 5 \cdot 10^{-4}$ and $I = 5 \cdot 10^{-5}$ for different couples of values of $\tilde{\sigma}_{yy}$ and  $\tilde{\dot{\gamma}}$. Coloured dots on both panels indicate the peak shear stress.}
    \label{fig:simple_shear_vol_press}
\end{figure}

In order to further explore this point, we compare two sets of simulations performed for a given value of $I$, however obtained using different couples of values of the applied strain-rate and imposed normal stress ($\tilde{\sigma}_{yy}, \tilde{\dot{\gamma}}$). This is done for $I$-values in the range $5\cdot10^{-5}\leq I \leq 5 \cdot 10^{-3}$, i.e. from the critical to the dense flow regime, see TABLE (\ref{tab:load_cond}). The results are shown on FIG. (\ref{fig:simple_shear_stress_press} and \ref{fig:simple_shear_vol_press}) and expressed in terms of the effective macroscopic friction $\mu= \tau / \sigma_{yy}$. We note that the differences between these two regimes, introduced earlier in this section, remain. Interestingly however, for a given value of $I$, the entire evolution of the macroscopic mechanical behavior, including the transient stages, appears independent on the way the loading is imposed (i.e. independent on the exact value of $\tilde{\sigma}_{yy}$ and $\tilde{\dot{\gamma}}$). The same is true for the evolution of the mean coordination number $Z$ and packing fraction $\phi$ (see FIG. \ref{fig:simple_shear_vol_press}). This suggests that the value of $I$ alone controls, at least at first order, the macroscopic behavior of the simulated granular media under simple shear over its entire loading history. This is quite unexpected, especially for the critical regime, as $I$ has been initially introduced to describe the competing effects of the imposed pressure versus shearing rate at the micro (or grain) scale from a \textit{local} point of view \cite{GDRMidi2004}, i.e. neglecting non-local effects \cite{Pouliquen2006,Forterre2008,Kamrin2012,Karimi2019} such as long-ranged elastic interactions, which are present especially in the critical regime and the early stages of deformation (see for instance FIG. \ref{fig:simple_shear_mm}a).

\subsection{\label{subsec:ela_prop} Elastic properties}

\begin{figure}[h]
    \begin{subfigure}{0.49\textwidth}
        \includegraphics[width=\textwidth]{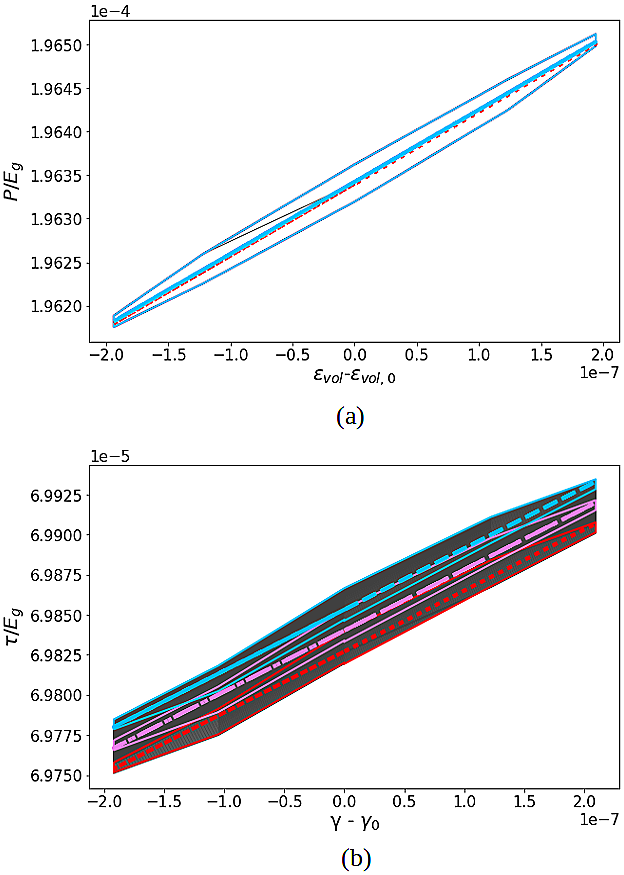}
    \end{subfigure}
    \caption{Evolution of (a) the pressure versus the imposed volumetric deformation and (b) the shear stress versus the shear strain, during an oscillatory test, for a $N = 10000$ grains system and for $I = 1 \cdot 10^{-4}$, corresponding to an imposed shear strain $\gamma(t=0) = 1.5 \cdot 10^{-3}$.}
    \label{fig:oscillatory_test}
\end{figure}

\begin{figure*}[!t]
    \centering
     \begin{subfigure}{0.99\textwidth}
        \includegraphics[width=\textwidth]{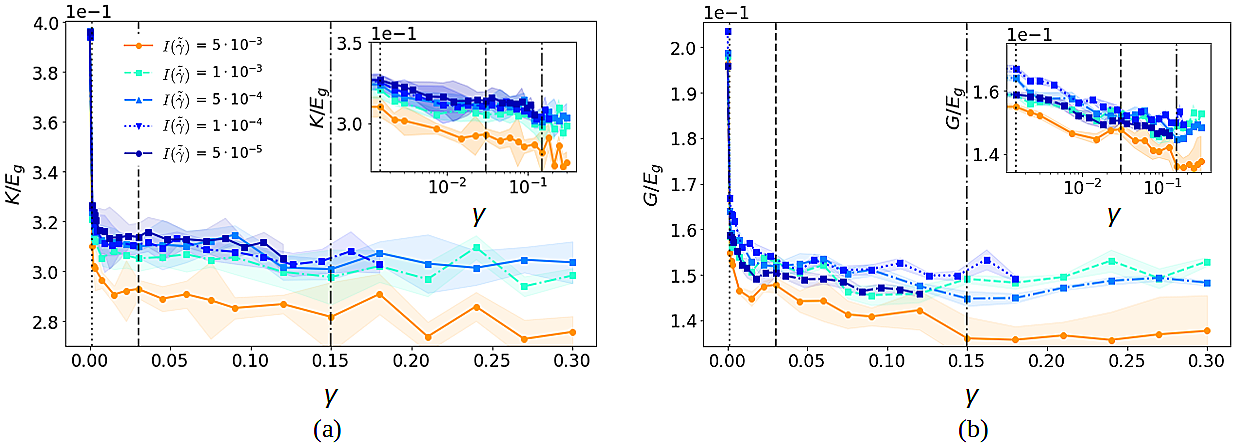}
    \end{subfigure}
    \caption{Evolution of (a) the bulk modulus $\tilde{K}$ and (b) the shear modulus $\tilde{G}$ as a function of the shear strain $\gamma$, during a simple shear experiments using $N = 10000$, $\tilde{\sigma}_{yy} = 1.0 \cdot 10^{-4}$ and four different values of $\tilde{\sigma}_{yy}$.}
    \label{fig:ela_prop}
\end{figure*}

\begin{figure*}[!t]
\centering
        \begin{subfigure}{0.99\textwidth}
                \includegraphics[width=\textwidth]{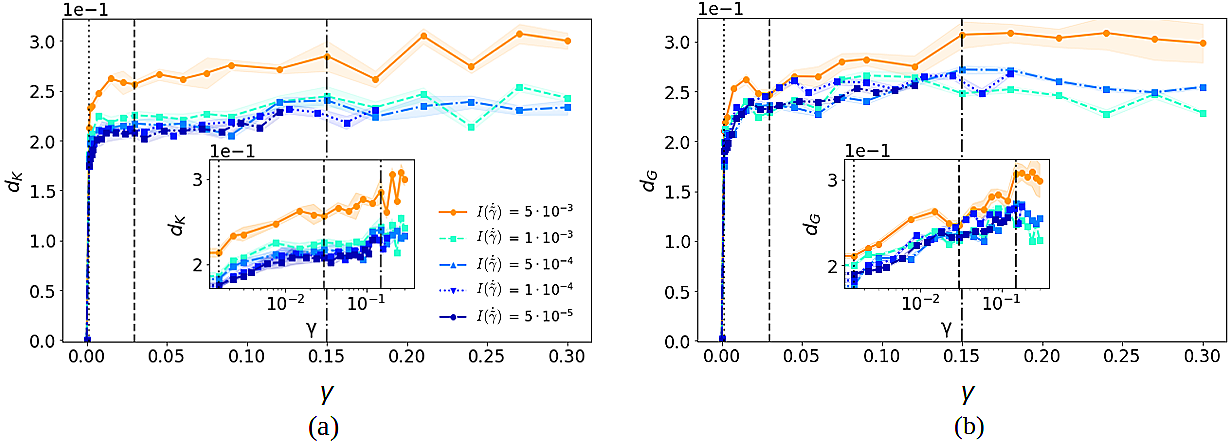}
        \end{subfigure}
    \caption{Evolution of the damage (a) $d_K$, associated with the degradation of $\tilde{K}$, and (b) $d_G$, associated with the degradation of $\tilde{G}$ as a function of the shear strain $\gamma$ during a simple shear experiments using $N = 10000$, $\tilde{\sigma}_{yy} = 1.0 \cdot 10^{-4}$ and four different values of $\tilde{\dot{\gamma}}$.}
    \label{fig:damage}
\end{figure*}
We estimate the elastic properties of our simulated granular media at $19$ different values of the imposed strain $\gamma$ during the simple shear experiment, via the small strain oscillatory tests described in section (\ref{sec:protocol}). 
FIG. \ref{fig:oscillatory_test}a and b shows respectively an example of stress-strain curve corresponding to a bi-axial compression and a pure shear oscillatory test for a sample loaded such that $I=1\cdot 10^{-4}$ and at a deformation of $\gamma=7.5\cdot 10^{-5}$, hence before the peak stress. 
For the biaxial compression test, a limit cycle is reached after the first oscillation. The small hysteresis suggests a negligible viscous dissipation within the granular assembly during the test. In the case of the pure shear test, a stress decrease of about $<1\%$ is observed over the 40 loading cycles. But the hysteresis as well as the slope of the main axis of the loop remain unchanged after the first cycle. The values of $\tilde{G}$ and $\tilde{K}$ reported below were estimated from the slope of the main axis of the loops \cite{Lewandowski2010} using the last 10 cycles of the oscillation tests. \\

The evolution of the elastic properties $\tilde{G}$ and $\tilde{K}$ as a function of the deformation $\gamma$ throughout the simple shear experiment is represented on FIG. \ref{fig:ela_prop} in the case of the following loading conditions: a normal stress $\tilde{\sigma}_{yy}=1.0 \cdot 10^{-5}$ and five values of prescribed $\tilde{\dot{\gamma}}$, such that $I$ ranges from the critical to the dense regime. It shows that both $\tilde{G}$ and $\tilde{K}$ initially decrease very fast, then much more slowly and proportionally to $\log(\gamma)$ for $\gamma \geq 10^{-3}$. 

For their small deformation, bi-axial compression numerical experiments on similar granular assemblies, Karimi et al. \cite{Karimi2019} reported an almost identical evolution of the shear modulus $G$ and of the average coordination number $Z$ up to the peak stress. It is important to note however that, in our simple shear experiments, we do not focus with great details on the small deformation regime that precedes the peak stress but instead extend the analysis to much larger strains. 
We note that the evolution of the estimated elastic properties (FIG. \ref{fig:ela_prop}) and of $Z$ (FIG. \ref{fig:simple_shear_vol}a) are similar over the entire loading history, with an initial sharp decrease followed by a much slower decay. However, beyond $\gamma=0.03$, $Z$ remains essentially constant while the elastic moduli continue to decrease. This indicates that the coordination number does not capture all the details of the evolution of the elastic properties, which can be degraded while $Z$ remains on average unchanged. 
At the macro scale of the granular assembly, we can interpret this elastic softening during shear deformation within the framework of continuum damage mechanics, however making the crucial distinction that the damage does not result from the breakage of the grains but from topological rearrangements. The effect of these rearrangements is to either eliminate grain contacts (decreasing $Z$) and/or modify force chains at constant $Z$, thereby deteriorating both the bulk and shear elastic stiffnesses. 
We further quantify a "shear" damage as $d_G = 1 - \tilde{G}/\tilde{G}_0$ and a "bulk" damage as $d_K = 1 - \tilde{K}/\tilde{K}_0$, with $\tilde{G_0}$ and $\tilde{K_0}$ the shear and bulk moduli estimated at $\gamma$ = $0$.
The evolution of $d_K$ and $d_G$ throughout the simple shear experiment is shown on FIG. \ref{fig:damage}. Much like the macroscopic shear stress and pressure (see section \ref{subsec:sh}), this damage evolution appears independent of the inertial number for $I\leq 10^{-3}$, i.e., within the critical regime. In the dense flow regime ($I=5\cdot10^{-3}$), the decrease in damage is initially more pronounced but evolves similarly as in the critical regime for larger strains. 

A macroscopic Poisson's coefficient can be computed for the 2D granular assembly from the measured elastic moduli as $\nu_{2D} = (\tilde{K} - \tilde{G}) / (\tilde{K} + \tilde{G})$. The values range between about $0.32 $ and $0.35$, which is close to the prescribed microscopic (i.e., grain) value. The evolution of $\nu_{2D}$ is characterized by a small increase in the early stage of deformation and a weak dependence on the applied strain rate (not shown) consistent with the granular assembly being slightly more compressible (lower $\nu_{2D}$) in the dense flow regime.

\begin{figure}[!h]
     \begin{subfigure}{0.49\textwidth}
        \includegraphics[width=\textwidth]{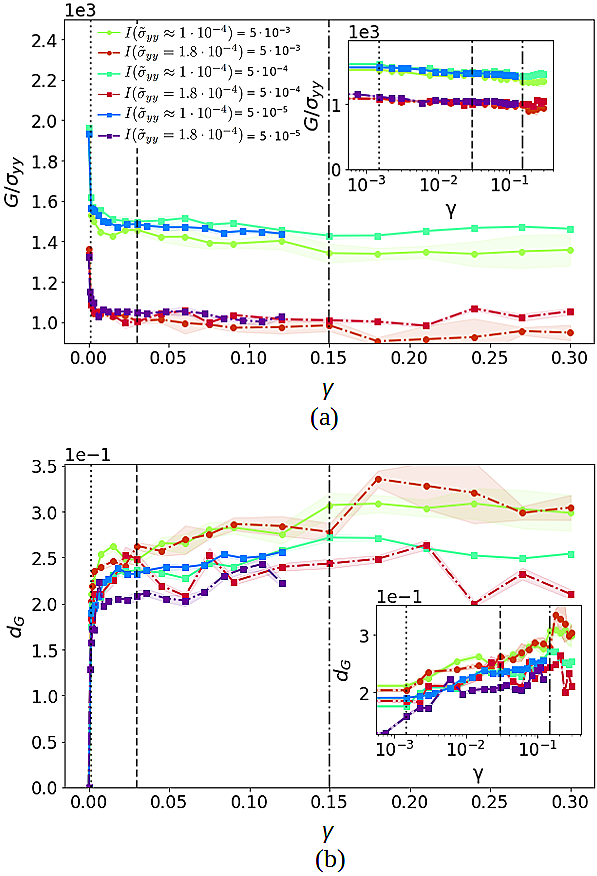}
    \end{subfigure}
    \caption{Evolution of (a) the shear modulus $\tilde{G}$ and (b) the shear damage $d_G$ as a function of the shear strain $\gamma$, in simple shear experiments using $N = 10000$, normalized by the applied normal stress $\tilde{\sigma}_{yy}$, and three different values of $I$ obtained for different couples of values of $\tilde{\sigma}_{yy}$ and  $\tilde{\dot{\gamma}}$. Coloured dots on both panels indicate the peak shear stress and pressure.}
    \label{fig:ela_dam_press}
\end{figure}
As done in Section \ref{subsec:sh}, we now raise the question of if and how the damage evolution depends on the imposed loading by exploring $I$ values obtained by prescribing different values of the normal stress $\tilde{\sigma}_{yy}$ and shear strain-rate $\tilde{\dot{\gamma}}$. The results, summarized in FIG. \ref{fig:ela_dam_press}a in the case of shear, show that the shear modulus $\tilde{G}$ is larger for a larger applied normal stress, in qualitative agreement with the effective medium theory of elasticity \cite{Walton1987}. However, when normalizing by the initial value of the shear modulus $\tilde{G_0}$, i.e., looking at the evolution of shear damage $d_G$ (see FIG. \ref{fig:ela_dam_press}b), these differences largely decrease. 
Very similar results are obtained for the bulk modulus $\tilde{K}$, bulk damage $d_K$ (not shown) and Poisson's ratio (not shown). 
Therefore, and surprisingly again, the value of the inertial number alone, not that of the pressure or strain rate forcings, appears to control the main features of the damage evolution from the critical regime to the transition to the dense regime, although this adimensional number has been proposed to characterize the rheology of granular media from a fluid point of view \cite{Jop2006}, that is, without accounting for the elastic behavior of their solid phase.

\subsection{\label{subsec:relax} Stress relaxation }

\begin{figure}[!h]
    \begin{subfigure}{0.5\textwidth}
        \includegraphics[width=\textwidth]{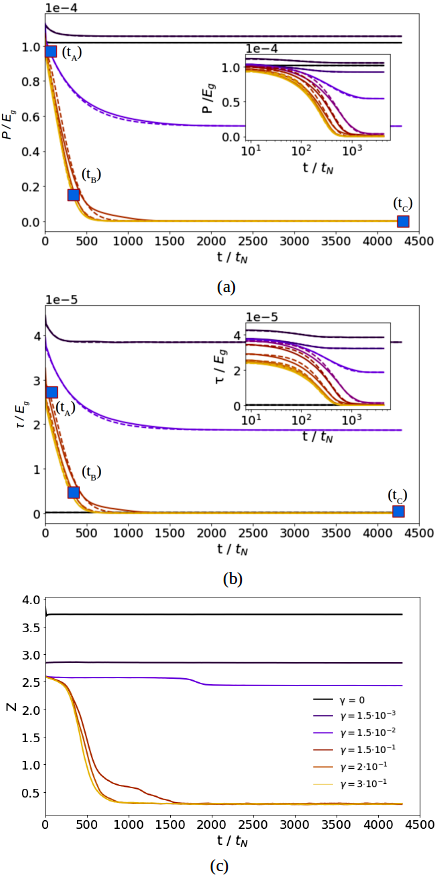}
    \end{subfigure}
    \caption{Temporal evolution of (a) the coordination number $Z$, (b) the macroscopic pressure $\tilde{P}$ and (c) the shear stress $\tilde{\tau}$ during relaxation tests performed with $N$ = $10000$ grains in the critical regime ($I = 5 \cdot 10^{-4}$). Solid lines on sub-figures (b) and (c) are the simulation results while the dashed lines are power-law fits (see Eq. \ref{eq:adim_pow_exp}). ($t_A$), ($t_B$) and ($t_C$) refer to the snapshots on FIG. \ref{fig:relax_mm}.}
    \label{fig:relax_stress}
\end{figure}

Relaxation tests show that when the imposed shear deformation is stopped and maintained constant, the simulated granular assembly relaxes, partly or completely (see FIG. \ref{fig:relax_stress} for a system in the critical regime with $I= 5 \cdot 10^{-4}$). 
A first important observation is that, when shearing is stopped after a small imposed deformation ($\gamma<0.01$), the system only partially relaxes towards a non-zero residual pressure $P_c$ and shear stress $\tau_c$. This indicates the occurrence of a plastic, or yield, stress. In comparison, when deforming the system beyond $\gamma=0.1$, the shear stress relaxes to zero and the pressure, almost to zero, consistent with a fluid.

As mentioned previously, Hartley and Behringer \cite{Hartley2003} reported a (partial) stress relaxation of their sheared granular assemblies of photo-elastic disks consistent with our results. They distinguished two regimes of relaxation: (i) a "fast" relaxation over the first $\sim 100$ seconds, followed by (ii) a very slow logarithmic relaxation over longer timescales. It has been argued that this second regime results from a slow aging of the frictional grain contacts \cite{Miksic2013}.
As this process is not implemented in our simulations, we do not expect to capture the associated logarithmic relaxation. In addition, if, for the relaxation experiments of \cite{Hartley2003} we would normalize the time by the vibrational timescale of the particles ($t_N$) as done here, this logarithmic regime would starts at an adimensional time of $\sim 10^5$, hence about two orders of magnitude larger than the duration of our own relaxation tests (see FIG. \ref{fig:relax_stress}). This confirms that the relaxation captured by our simulations most likely corresponds to the first, fast relaxation regime.

\begin{figure}[!t]
    \begin{subfigure}{0.49\textwidth}
        \includegraphics[width=\textwidth]{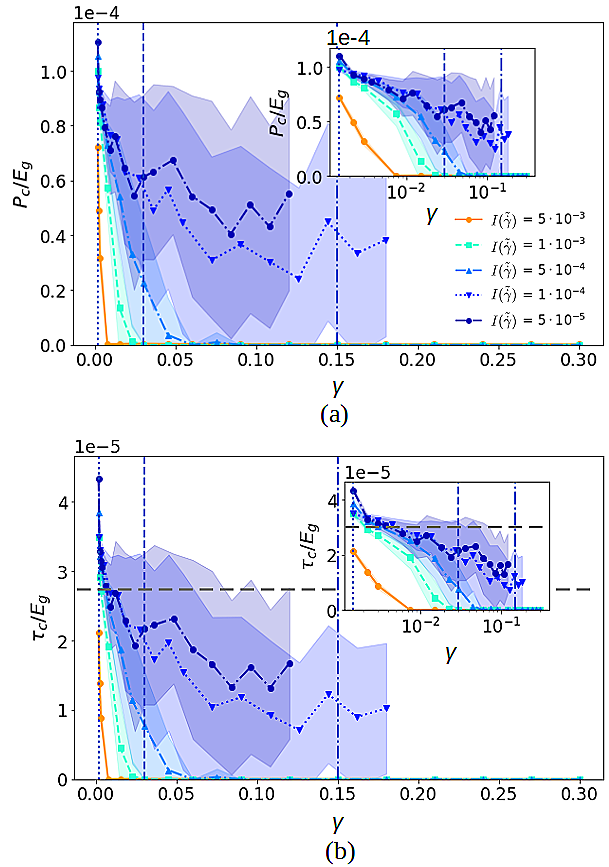}
    \end{subfigure}
    \caption{Evolution of (a) residual pressure $\tilde{P}_c$ and (b) residual shear stress $\tilde{\tau}_c$ as a function of the shear strain $\gamma$ for a $N = 10000$ grains sample under an imposed compression $\tilde{\sigma}_{yy} = 1 \cdot 10^{-4}$. The value predicted by $\mu(I)$ rheology $\mu_1 \cdot \tilde{\sigma}_{yy} = 2.8 \cdot 10^{-5}$ is shown (solid grey dash line).}
    \label{fig:relax_stress_vs_gamma}
\end{figure}

 \begin{figure*}[!t]
        \begin{subfigure}{0.98\textwidth}
        \includegraphics[width=\textwidth]{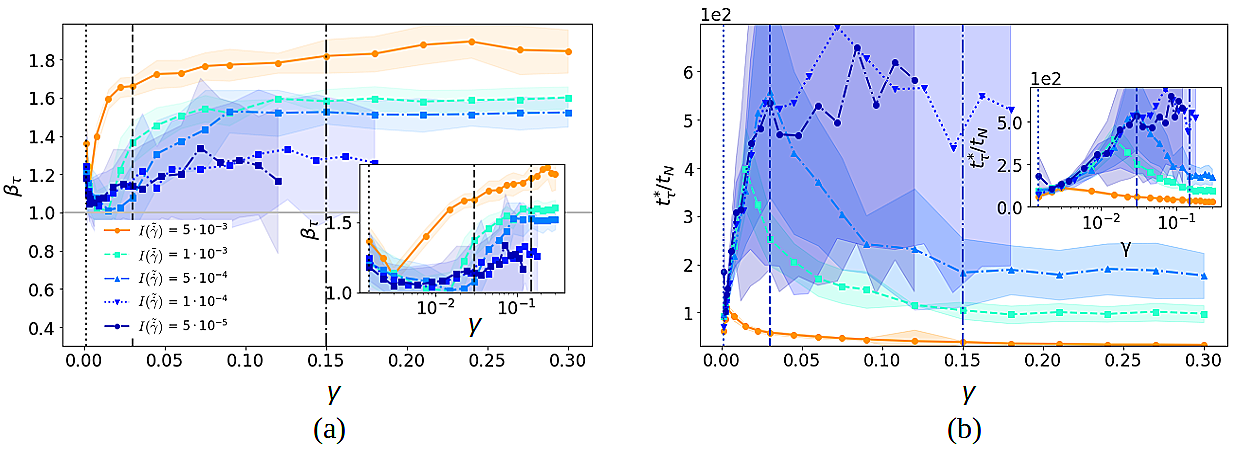}
    \end{subfigure}
    \caption{Evolution of (a) the coefficient of shear stress relaxation $\beta_{\tau}$ and (b) the characteristic time of relaxation of the shear stress $\tilde{t}_{\tau}^*$ as a function of the shear strain $\gamma$ in simple shear experiments using $N = 10000$, different values of $I$ and an imposed compression $\tilde{\sigma}_{yy} = 1 \cdot 10^{-4}$.}
    \label{fig:relax_time_vs_gamma}
\end{figure*}
The evolutions of the residual pressure $\tilde{P}_c$ and shear stress $\tilde{\tau}_c$ with the applied shear deformation $\gamma$ are shown respectively on FIG. \ref{fig:relax_stress_vs_gamma}a and \ref{fig:relax_stress_vs_gamma}b for different values of $I$ ranging from the critical to the dense flow regime. Here, $I$ was set by varying the applied strain rate, $\tilde{\dot{\gamma}}$, at a fixed value of $\tilde{\sigma}_{yy}$.

These residual stresses decay with $\gamma$ in all cases. However, for very slow shear deformation rates, i.e. $I\leq10^{-4}$, they do not vanish, even in the stationary stage ($\gamma > 0.15$, described in section \ref{subsec:sh}). For fast deformation rates, i.e., $I>10^{-4}$, they do vanish and that, earlier with increasing $I$, which supports a fluid behavior passed the first three and transient stages. 
This is in qualitative agreement with the $\mu(I)$ framework, which describes a transition from a solid (in the critical regime) to a fluid (in the dense flow regime). 

We now turn our attention to the form of this relaxation, which we fit with the following generic function: 
\begin{equation}
    \frac{\sigma(t) - \sigma_c}{\sigma_0 - \sigma_c} = \exp \left[-\left(\frac{t}{\iota^*}\right)^{\beta}\right],
    \label{eq:adim_pow_exp}
\end{equation}
where $\sigma(t)$ is the stress (either the pressure or the shear stress), $\sigma_c$ the residual stress at the end of the relaxation test, $\sigma_0$ the initial stress at $t = 0$, $\iota^*$ a timescale and $\beta$ an exponent. Integrating this function in time gives a characteristic time $t^*$:
\begin{equation}
    t^*  = \int^{\infty}_{0} \exp{\left(-\frac{t}{\iota^*}\right)^{\beta}} dt = \frac{\iota^*}{\beta} \cdot \Gamma\left(\frac{1}{\beta}\right),
    \label{eq:mean_relax_t}
\end{equation}
where $\Gamma(x)$ is the Gamma function of parameter $x$.
Eq. (\ref{eq:mean_relax_t}) nicely fits our simulations results (see FIG. \ref{fig:relax_stress}), thereby allowing to estimate $\beta$ and $t^*$. 

FIG. \ref{fig:relax_time_vs_gamma} shows the evolution of $\beta$ and $t^*$ with the applied shear deformation $\gamma$, for different values of $I$ ranging from the critical to the dense flow regime, in the case of the relaxation of the shear stress. Here, again, $I$ was set by varying the applied strain rate, $\tilde{\dot{\gamma}}$, at a fixed value of $\tilde{\sigma}_{yy}$.
Very similar results were obtained for the relaxation of the pressure (not shown). 
A first observation is that these relaxations are always compressed ($\beta > 1$), and characterized by a decrease of $\beta$ during the early stage of the deformation, followed by an asymptotic increase at later stages. Making an analogy with glasses \cite{Trachenko2021}, such "fast" compressed relaxations could be the signature of avalanches of topological rearrangements that are triggered by elastic interactions within the system. Examining the microscopic evolution of the system supports this interpretation (see below). Surprisingly, however, the compressed character of the relaxation (quantified by $\beta$) increases with increasing inertial number $I$ (see FIG. \ref{fig:relax_time_vs_gamma}), i.e. for systems which are more "fluid" in terms of residual stresses. Second, for all $I$, the characteristic time $t^*$ is characterized by a rapid increase in the early stages of the deformation, followed by a slower and asymptotic decrease at later stages. Its value increases with decreasing $I$, indicating that the relaxation is further delayed for slowly sheared systems. We mention here that, for a compressed exponential, the value of $t^*$ as defined in Eq. (\ref{eq:mean_relax_t}) roughly corresponds to the time at which the relaxation occurs at the fastest rate.  

\begin{figure}[!h]
    \begin{subfigure}{0.5\textwidth}
        \centering
        \includegraphics[width=\textwidth]{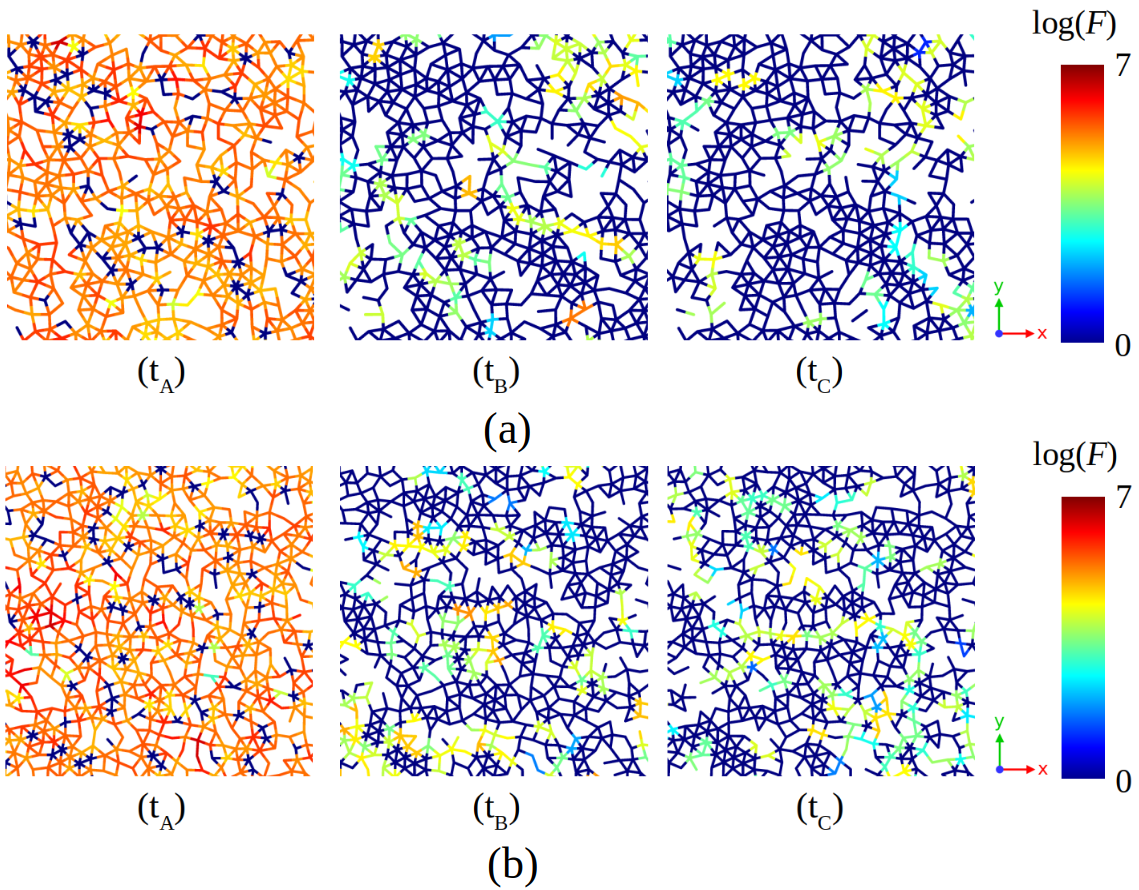}
    \end{subfigure}
    
    \caption{Microscopic evolution of the force chains for a system with (a) $I = 5 \cdot 10^{-3}$ and (b) $I = 5 \cdot 10^{-4}$ and $N = 10000$. Snapshots are taken during the relaxation of the system at times $t_A/t_N$ = $1 \cdot 10^2$, $t_B/t_N$ = $2 \cdot 10^2$ and $t_C/t_N$ = $4.2 \cdot 10^3$.}
    \label{fig:relax_mm}
\end{figure}

\begin{figure}[!h]
    \centering
    \begin{subfigure}{0.5\textwidth}
        \includegraphics[width=\textwidth]{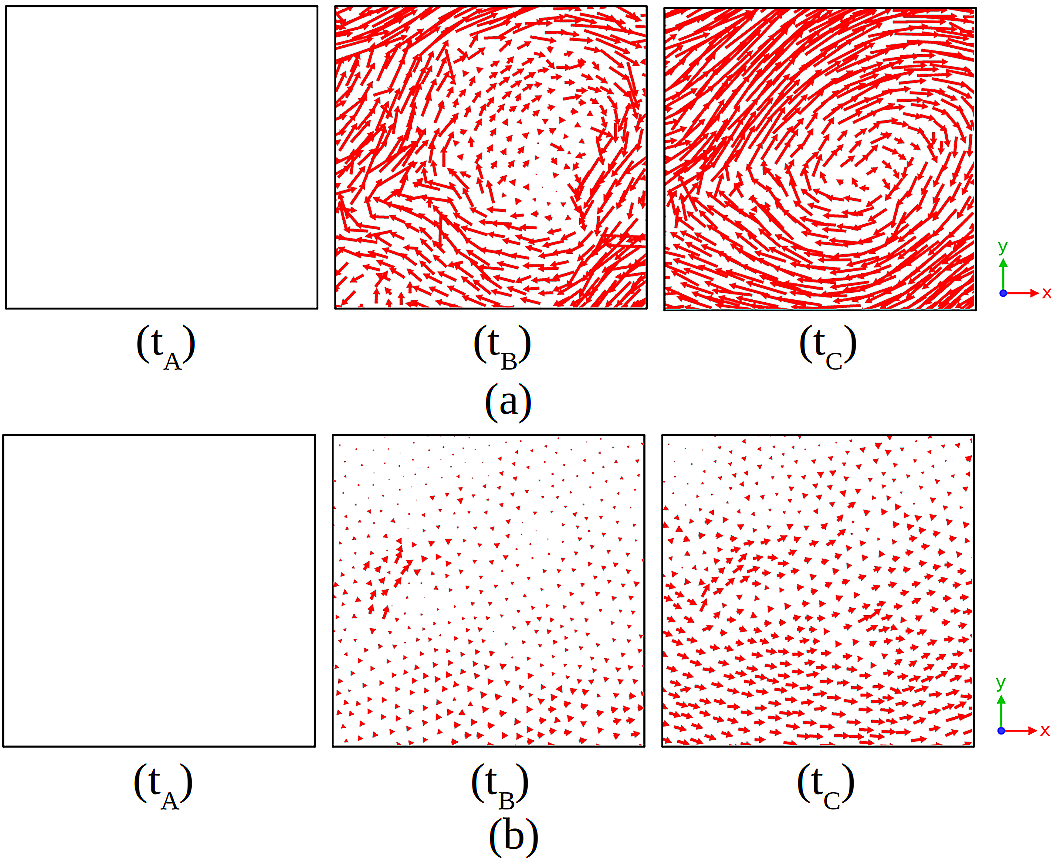}
    \end{subfigure}
    \caption{Non-affine cumulative displacement monitored during relaxation of the granular assemblies at (a) $I = 5 \cdot 10^{-3}$, (b) $I$ = $5 \cdot 10^{-4}$ along the simulation for a $N = 10,000$ grains sample. Micro-mechanical snap shots where taken for the relaxation of the system at the times ($t_A$), ($t_B$) and ($t_C$) such as $t_A/t_N$ = $1 \cdot 10^2$, $t_B/t_N$ = $2 \cdot 10^2$ and $t_C/t_N$ = $4.2 \cdot 10^3$. Magnification by a factor $\times 5 \cdot 10^{5}$ at $t_a$ and $t_b$ and $\times 5 \cdot 10^4$ at $t_c$. Similar vortices are observed throughout shearing (not shown).}
    \label{fig:vortexe_mm}
\end{figure}

Observation of the evolution of force chains measured at the end of shearing ($\gamma = 0.3$) further argue for avalanches of local rearrangements. FIG. \ref{fig:relax_stress}a and \ref{fig:relax_mm}a indeed show that during relaxation most of the force chains collapse between adimensional times $t_A\sim 100$ and $t_B \sim 1000$ and that, both in the dense flow ($I= 5 \cdot 10^{-3}$) and critical ($I = 5 \cdot 10^{-4}$) regime. During this same time interval, the average coordination number, represented in FIG.\ref{fig:relax_stress}c in the case of $I = 5 \cdot 10^{-4}$, decreases rapidly to zero. 
The non-affine grain cumulative displacement fields recorded during the relaxation of the systems previously sheared up to $\gamma=0.3$ under different strain-rates (FIG. \ref{fig:vortexe_mm}) show that these relaxations occur through spatially-organized rearrangements taking the form of well-defined vortices for samples sheared under large $I$-values.

\begin{figure}[!h]
    \begin{subfigure}{0.5\textwidth}
        \includegraphics[width=\textwidth]{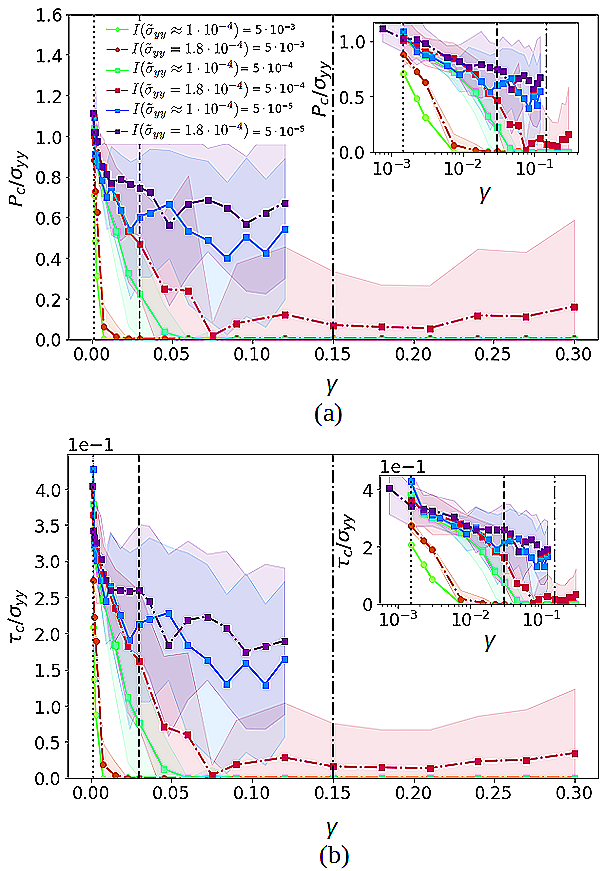}
    \end{subfigure}
    \caption{Evolution of (a) the residual pressure $\tilde{P}_c$ and (b) the residual shear stress $\tilde{\tau}_c$ as a function of the shear strain $\gamma$ in simple shear experiments using $N = 10000$, normalized by the applied normal stress $\tilde{\sigma}_{yy}$, and three different values of $I$ obtained for different couples of values of $\tilde{\sigma}_{yy}$ and $\tilde{\dot{\gamma}}$.}
    \label{fig:relax_time_vs_gamma_press}
\end{figure}

As for the macroscopic behavior and evolution of the elastic properties, we conclude this section by comparing the relaxation of the stresses in systems with the same inertial number $I$ but deformed by prescribing different values of the normal stress $\tilde{\sigma}_{yy}$ and shear strain-rate $\tilde{\sigma}_{yy}$ (see FIG. \ref{fig:relax_time_vs_gamma_press}a and \ref{fig:relax_time_vs_gamma_press}b). Residual pressures $\tilde{P_c}$ and shear stresses $\tilde{\tau}_c$ are larger for systems sheared under a larger normal stress $\tilde{\sigma}_{yy}$ (not shown). However, after normalization by $\tilde{\sigma}_{yy}$, the differences observed between systems with the same $I$ become small compared to that observed between systems with a different $I$. This suggests once again that the inertial number is a controlling factor of the mechanical behavior of frictional granular media, including properties (damage and relaxation) that are beyond the scope of the $\mu(I)$ rheology.

 \begin{figure*}[!t]
        \centering
        \begin{subfigure}{0.99\textwidth}
        \includegraphics[width=\textwidth]{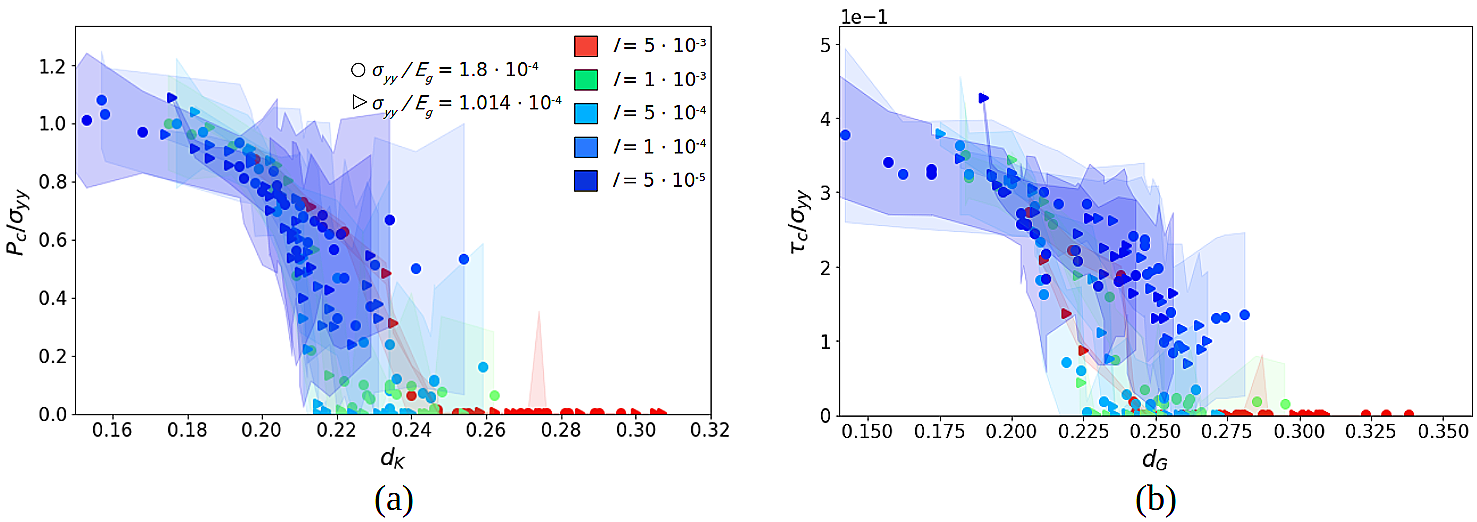}
    \end{subfigure}
        \caption{(a) Residual pressure $P_c/\sigma_{yy}$ as function of the bulk damage $d_K$ and (b) residual shear stress $\tau_c/\sigma_{yy}$ as a function of the shear damage $d_G$ for systems with $N = 10000$, normalized by the applied normal stress $\tilde{\sigma}_{yy}$, and different values of the inertial numbers $I$ obtained for different couples of values of $\tilde{\sigma}_{yy}$ and $\tilde{\dot{\gamma}}$.}
    \label{fig:sigma_c_vs_dam}
\end{figure*}

\section{\label{sec:discussion} Discussion}

\subsection{\label{subsec:inertial_number} $I$-dependent behavior}

Previous studies have shown that a single dimensionless number, the inertial number $I$, describing the relative importance of inertia and pressure at the scale of the grains, plays a key role in the flow properties of frictional, dense granular assemblies \cite{GDRMidi2004}. 
These works have considered perfectly rigid spherical grains, hence not taken into account elastic properties and interactions, and a clear separation of scales between the (i) the system size, (ii) the scale of individual grains and grain-grain contacts and (iii) the microscopic scales at which the contacts are established, including the roughness of grain surfaces. The associated $\mu(I)$ rheology (see Eq. \ref{eq:mu_of_I}) was established assuming that the microscopic scale does not influence the flow properties, as correlation lengths within the flow are considered not to extend significantly beyond the scale of individual grains.

In the present work, we have explored the mechanical behavior of a simulated 2D frictional granular medium under conditions for which these assumptions are not expected to hold, that is, in the critical regime and up to the dense flow regime (FIG. \ref{fig:mu_I_results}). Although we have verified that our model does comply to the $\mu(I)$ prediction in the regime in which it is expected to hold, the dense flow regime (section \ref{muI}), the emerging non-affine velocity fields (see FIG. \ref{fig:vortexe_mm}) is clearly organized in vortices with correlation lengths that are much larger than the grain size, in agreement with previous observations in the dense \cite{Rognon2015} and critical regime \cite{Radjai2002,Combe2013}. 
The granular assemblies were studied under an imposed normal stress that satisfies the small perturbations assumption of \cite{Mindlin1949}, with an applied pressure remaining below $\tilde{P} <  2.2 \cdot 10^{-4}$ \cite{daCruz2005}.
We have shown that our simulated granular medium presents a finite macroscopic elastic stiffness, which partially deteriorates during shear deformation (section \ref{subsec:ela_prop}) due to topological re-arrangements of the grains.
Additionally, we argued on section \ref{subsec:relax} that the compressed relaxations result from avalanches of these topological rearrangements, a mechanism necessarily involving long-ranged interactions as well. 

These disparities between the $\mu(I)$ framework and our simulations imply a limited role of the inertial number $I$ in predicting the mechanical behavior of our simulated granular medium. 
Nevertheless, we have shown that, for a given value of $I$, its behavior under simple shear as well as its damage and relaxation properties are independent of the details of the forcing, that is, on the specific value of the applied pressure and shear strain rate.
These unexpected results suggest that the inertial number $I$ captures some aspects of these processes as well as of the macroscopic behavior of frictional granular media in the critical regime. 
However, as mentioned in section \ref{sec:shear_test} and further discussed below, this does not entail that the $\mu(I)$ \textit{rheology} describes fully the mechanical behavior of our simulated granular medium and, probably, of frictional granular media in general in this critical regime.

\subsection{\label{subsec:prop_corel} Relation between the damage and relaxation properties}
First, it is important to recall that the $\mu(I)$ rheology, by construction (see Eq.\ref{eq:mu_of_I}), cannot describe the transient stages of shear deformation described in section \ref{subsec:sh}, but only the stationary, asymptotic behavior reached at large shear strains. 
It does predict the emergence of two different phases, as it represents the medium as visco-plastic. Below a shear threshold $\mu_1P$, it predicts a perfectly rigid solid behavior (again, elasticity is not considered). Above this threshold, it predicts a viscous fluid behavior, characterized by a pressure- and strain-rate-dependent viscosity (see section \ref{sec:intro}). 
From this description, we should expect the following relaxation behavior :
(i) an absence of significant stress relaxation of the system in the critical regime, i.e. for $I\lesssim10^{-3}$, as the shear stress is already at the threshold $\mu_1 \cdot P$
(ii) a relaxation down to a residual shear stress $\tau_c=\mu_1 \cdot P$ for systems in the dense flow regime.

The results reported in section \ref{subsec:relax} reveal a more complex scenario. 
Even for very slowly driven systems ($I\leq10^{-4}$), the shear stress partly relaxes to a residual shear stress $\tau_c/\sigma_{yy}$ slightly lower than expected from a direct interpretation of the $\mu(I)$ visco-plastic rheology, i.e. $\tilde{\tau}_c\sim 0.1-0.2 < \mu_1 \cdot \tilde{\sigma}_{yy}= 0.28$, see FIG.(\ref{fig:relax_stress_vs_gamma}b). 
From $I=5\cdot10^{-4}$ and beyond, i.e. the dense flow regime, the shear stress relaxes completely ($\tilde{\tau}_c\rightarrow 0$), an unexpected behavior for a visco-plastic fluid.
In addition, for a given system, the relaxation evolves significantly in the course of shear deformation. 
Therefore, although on the point view of its frictional behavior in the stationary regime, our simulated granular medium agrees with a $\mu(I)$ rheology (FIG. \ref{fig:mu_I_results}), it does not behave as a visco-plastic fluid, but rather as a more complex visco-elasto-plastic fluid, with damage-dependent elastic, relaxation and plastic properties. 

Indeed, we have shown that it is characterized by a non-zero bulk and shear elastic stiffnesses over the entire course of shearing deformation. Both elastic modulii $\tilde{K}$ and $\tilde{G}$ decrease with increasing $\gamma$, suggesting a progressive damage of the system. This damage quantifies at the macroscopic scale the magnitude and spatial organization of the microscopic force chains \cite{Karimi2019}. 

In turn, we expect the relaxation properties to strongly depend on the topology of these chains of force, thus raising the question of a possible relation between damage and relaxation. 
We investigate this relation by plotting the residual pressure $\tilde{P}_c$ as a function of the bulk damage parameter $d_K$, and the residual shear stress $\tilde{\tau}_c$ as a function of the shear damage parameter $d_G$.
FIG.\ref{fig:sigma_c_vs_dam}a and \ref{fig:sigma_c_vs_dam}b compile this data for all of the relaxation tests, performed at different cumulative shear deformation $\gamma$, and all of the explored values of $I$, obtained for different couples of values of applied $\tilde{\sigma}_{yy}$ and $\tilde{\dot{\gamma}}$. A general pattern emerges, which is similar in both cases: beyond a damage value of $d_{K,G}\simeq0.23=d_c$, the residual pressure and shear stress vanish, a relaxation behavior compatible with a viscous fluid, while for $d < d_c$, they increase sharply towards a seemingly asymptotic value at $d\rightarrow0$, indicating that a plastic component of the rheology remains. This characterizes a phase transition from a solid- to fluid-like behavior, which is controlled by a critical damage value $d_c$.

\section{\label{sec:conclusion} Conclusions}

Our numerical simulations of the simple shear deformation of athermal, dry, frictional granular media in the critical and dense flow regimes show that the mechanical behavior of these media is more complex than that of a viscous-plastic fluid, as it depends not only on the loading conditions, through the inertial number $I$, but on the loading history (i.e., the cumulated shear deformation) as well. Within the range of values of the inertial number explored here, our simulated granular medium has a significant elastic stiffness, which partially deteriorate during the shear deformation: a process that can be quantified via a damage parameter. The meaning of this damage is fundamentally different than in continuum mechanics, as it is not related to fracturing processes but to topological re-arrangements of the grains. Our simulation results suggest that it is a first-order controlling parameter of the rheology of dry, frictional granular media. Indeed, beyond a threshold value, the medium behaves as a viscous material, in which stresses dissipate completely when deformation is maintained fixed, while below this threshold, a residual stress remains. The elasto-visco-plastic behavior in this later limit is non-trivial, as the temporal evolution of the relaxation follows a compressed exponential, which highlights the importance of elastic and long-range grain-grain interactions. The formulation of an associated damage-dependent, visco-elasto-plastic constitutive law for frictional granular media is an ongoing task, which could find applications in a number of geophysical and more general contexts.

\begin{acknowledgments}

We thank Kamran Karimi and Agnieszka Herman for their help with the LAMMPS software and fruitful discussions on the development of the code. All authors are supported through the Scale Aware Sea Ice Project (SASIP). SASIP is supported by SchmidtSciences, LLC. \\ 

\end{acknowledgments}

\newpage
\nocite{*}
\bibliography{sorsamp} 

\end{document}